\documentclass[twocolumn]{aastex631} 
\usepackage{amsmath}
\usepackage{amssymb}
\usepackage{graphicx}
\usepackage{mathtools}
\usepackage{amsmath,esint}

\shorttitle{Interstellar Objects}
\shortauthors{Hoover, Seligman and Payne}
\graphicspath{{./}{figures/}}

\begin{document}

\title{The Population of Interstellar Objects Detectable with the LSST and Accessible for \textit{In Situ} Rendezvous with Various Mission Designs  }

\correspondingauthor{Devin J. Hoover}
\email{djhoover99@uchicago.edu}

\author[0000-0002-7783-6397]{Devin J. Hoover}
\affil{Dept. of Astronomy and Astrophysics, University of Chicago, Chicago, IL 60637}

\author[0000-0002-0726-6480]{Darryl  Z. Seligman}
\affil{Dept. of the Geophysical Sciences, University of Chicago, Chicago, IL 60637}

\author[0000-0001-5133-6303]{Matthew J. Payne}
\affiliation{Harvard-Smithsonian Center for Astrophysics, 60 Garden St., MS 51, Cambridge, MA 02138, USA}


\begin{abstract}
The recently discovered population of interstellar objects  presents us with the opportunity to characterize  material from extrasolar planetary and stellar systems up close. The forthcoming Vera C. Rubin Observatory Legacy Survey of Space and Time (LSST) will provide an unprecedented increase in  sensitivity to these  objects compared to the capabilities of currently operational observational facilities. We generate a synthetic population of `Oumuamua-like objects  drawn from their galactic kinematics, and identify the distribution of impact parameters, eccentricities, hyperbolic velocities  and sky locations of  objects detectable with the LSST, assuming no cometary activity. This  population is characterized by a clustering of trajectories in the direction of the solar apex and anti-apex, centered at orbital inclinations of $\sim90^\circ$. We identify the ecliptic or solar apex as the optimal sky locations to search for future interstellar objects as a function of survey limiting magnitude. Moreover, we identify the trajectories  of detectable  objects that will be reachable for  \textit{in-situ} rendezvous with a dedicated mission with the capabilities of the forthcoming \textit{Comet Interceptor} or proposed \textit{Bridge} concept. By scaling our fractional population statistics with the inferred spatial number density,  we estimate that the LSST will detect of order $\sim 15$ interstellar objects over the course of its $\sim10$ year observational campaign.  Furthermore, we find that there should be $\sim1-3$ and $\sim0.0007-0.001$ reachable targets for missions with propulsion capabilities comparable to \textit{Bridge} and \textit{Comet Interceptor}, respectively.  These numbers are lower limits, and will be readily updateable when the number density and size frequency distribution of interstellar objects are better constrained.
\end{abstract}




\section{Introduction} \label{sec:intro}

The recent discoveries of the first two interstellar objects (ISOs), 1I/2017 U1 (`Oumuamua) and  C/2019 2I (Borisov), imply that a galactic-wide population of roughly $10^{26}$ similar bodies exist, with spatial number densities of order $n_{o}\sim1-2\times 10^{-1}\,$\textsc{au}$^{-3}$ \citep{Trilling2017,Laughlin2017,Do2018,moro2019a,moro2018,Levine2021}. These number densities have been estimated by incorporating the detailed detection capabilities of current observational facilities. Forthcoming surveys, such as the Vera C. Rubin Observatory Legacy Survey of Space and Time (LSST),  will provide unprecedented completeness for  populations of minor bodies in the Solar System \citep{jones2009lsst,Ivezic2019}. It has  been estimated that the LSST will detect between $1-10$ interstellar objects every year \citep{Engelhardt2014,Cook2016,Trilling2017,Seligman2018}. It is important to note that the detection frequency depends sensitively on the spatial distribution, size-frequency distribution and levels of cometary activity of these objects. The future detections and characterizations  of these objects will yield insights into the physical mechanisms at work in  extra-solar planetary systems and star formation regions throughout the galaxy.

The first interstellar object, `Oumuamua, defied all expectations for an interstellar comet. It exhibited a distinct lack of dust and typical cometary volatiles \citep{Meech2017,jewitt2017observation,trilling2018spitzer}, an extreme $6:6:1$ aspect ratio \citep{Knight2017,Drahus2017,Bolin2017,McNeill2018,Belton2018,Mashchenko2019}, a reddened reflection spectrum \citep{masiero2017spectrum,Fitzsimmons2017,Bannister2017,Ye2017}, and a non-gravitational acceleration \citep{Micheli2018}. Curiously, it also exhibited a surprisingly low incoming velocity with respect to the Local Standard of Rest (LSR) in the direction of the solar apex \citep{mamajek2017}, implying a young $<40$Myr age \citep{Fernandes2018,hallatt2020dynamics,Hsieh2021}. 

This peculiar combination of unique physical properties has led to a variety of theories describing the provenance of `Oumuamua. The non-gravitational acceleration has been postulated to be driven by radiation pressure \citep{Micheli2018}, which could be explained if `Oumuamua was an ultra low-density fractal aggregate \citep{moro2019fractal,luu2020oumuamua,Sekanina2019b}, or a millimeter-thin artificial light sail \citep{bialy2018radiation}. If the non-gravitational acceleration was powered by cometary outgassing \citep{Micheli2018,seligman2019acceleration}, it has been demonstrated that the energetics are consistent with the sublimation of H$_2$ \citep{fuglistaler2015solid,seligman2020h2,Levine2021_h2}, N$_2$ \citep{jackson20211i,desch20211i}, or CO \citep{Seligman2021}. Recently, \citet{Flekkoy2022} calculated infrared and optical spectral signatures that would differentiate between the various proposed scenarios.

In stark contrast to `Oumuamua, the interstellar comet Borisov displayed physical characteristics that are broadly consistent with the census of cometary bodies in the Solar System. It had  dusty cometary activity \citep{Jewitt2019b,Bolin2019,Fitzsimmons:2019,Ye:2019,McKay2020,Guzik:2020,Hui2020,Kim2020,Cremonese2020,yang2021}, and early observations confirmed the presence of carbon and nitrogen based species in the outflow \citep{Opitom:2019-borisov, Kareta:2019, Lin:2019,Bannister2020,Xing2020,Aravind2021}. It was particularly enriched in CO \citep{Bodewits2020, Cordiner2020}, which could be explained if it  formed in the outer regions of a protoplanetary disk \citep{Price2021}. It exhibited a brightening event \citep{Drahus2020ATel} and subsequent breakup in the spring of 2020 \citep{Jewitt2020:BorisovBreakup, Jewitt2020ATel, Bolin2020ATel,Zhang2020ATel}, which was likely due to seasonal effects \citep{Kim2020}.  Recently, \citet{Guzik2021} reported a spectroscopic detection of atomic nickel vapor in the  coma. \citet{Bagnulo2021} presented polarimetric observations  that demonstrated that the dust coma had abnormally high polarization compared to Solar System comets.  Astrometric analyses of the non-gravitational acceleration are consistent with observed levels of outgassing \citep{Hui2020,delafuente2020,Manzini2020}. 

Future detections and subsequent characterization of interstellar objects will be a welcome development. They should contextualize and/or explain the discrepancy in the observed characteristics of the first two discovered members of the population. In addition to ground and space-based observations of these future objects, \textit{in situ} measurements would provide an  opportunity to characterize their composition and physical properties in detail \citep{Seligman2018}.  ESA’s  \textit{Comet Interceptor} mission \citep{jones2019,Sanchez2021}  or the NASA concept study \textit{Bridge} \citep{Moore2021} may be well-positioned to provide these observations.

In this paper, we investigate the population of interstellar objects that will be detected by the LSST, with a focus on objects that could serve as realistic targets for interception missions. In \S \ref{section:Sample}, we describe the galactic distribution of interstellar objects from which we draw our simulated targets. In \S \ref{section:MC_integration}, we describe the numerical simulations that we perform in order to evaluate the trajectories and distances from the Earth that each of these objects attain. In \S \ref{section:Detectability} and \S \ref{section:Intercept}, we present the distribution of trajectories and orbital elements for objects that are (i) detectable by the LSST and (ii)  realistic targets for an interception mission. We also  provide an up-to-date estimation for the number of objects that will be detected for both of these populations. In \S \ref{section:Conclusions}, we summarize our calculations and conclude.

\section{The Galactic Population of Interstellar Objects} \label{section:Sample}

In this section, we describe the methodology with which we simulate a galactic population of interstellar objects. We assume that the population is dynamically relaxed  into the same  kinematic distribution as local stars via scattering events off of giant molecular clouds and dark matter substructure  \citep{Seligman2018}. Since this population  consists of $\sim10^{26}$ objects, it represents an excellent realization of the fine-grain assumption in the collisionless Boltzmann equation. 

We assume that the interstellar objects within the solar neighborhood follow a Schwarzschild velocity distribution with respect to the LSR, with velocity dispersions  $\sigma_{R}$, $\sigma_{\phi}$, $\sigma_{z}$. The right-handed galactic coordinate system is oriented by the vertex deviation, $v_{d}$, for the primary eigenvector of the ellipsoidal distribution function. The spatial number density of ISOs, $n({\bf \Delta v})$,  in a specific volume of phase space ${\bf \Delta v} = (\Delta v_1,\Delta v_2, \Delta v_3)$, is given by,

\begin{equation}\label{eq:Schwarzschild}
    n({\bf \Delta v}) =\frac{n}{8}\, \bigg(\, \prod_{i=1}^3 \textrm{erf}\big(\frac{v_i}{\sqrt{2}\sigma_i}\big)|_{v_i-\Delta v_i}^{v_i+\Delta v_i} \bigg) \, .
\end{equation}
Note that in this equation the numbered subscript $i$ is used to indicate the orthogonal components of $\sigma$ in the order previously indicated.

We adopt the values of $\sigma_{R}$, $\sigma_{\phi}$, $\sigma_{z}$, and $v_{d}$ for G stars \citep{Binney1998,Seligman2018}.  It is important to note that different stellar populations correspond to different typical ages and  velocity dispersions, as summarized in Table \ref{schwarzschild_parameters}. The typical ages of interstellar objects is somewhat unconstrained.  We verified that the results of our simulations are not sensitively dependant on the spectral type that we use. It would be informative to perform the calculations presented in \S \ref{section:Detectability} and \S \ref{section:Intercept} for  spectral classes representative of older and younger populations, but this is outside of the scope of this paper.

Gravitational focusing preferentially increases the flux of interstellar objects with low heliocentric velocities by a factor $\xi$, which we approximate as,
\begin{equation}\label{focusing_prob}
    \xi\approx 1 + (\frac{42 km/s}{v_{\infty} + \beta})^2\,.
\end{equation}
Here, $v_{\infty}$ is the hyperbolic excess velocity with respect to the Sun and $\beta$ is an artificial smoothing factor that we set to $\beta=$2 km/s to avoid the hyperbolic case where $v_{\infty}=0$ \citep{Seligman2018}.   We  compute the probability that an interstellar object enters the Solar System with a given velocity using Equations \ref{eq:Schwarzschild} and \ref{focusing_prob}.

We construct a three-dimensional galactic velocity grid by generating 100 evenly-spaced values for each component of heliocentric velocity between -100 and 100 km/s.  We calculate the probability that an interstellar object will enter the Solar System  for each  grid cell and corresponding velocity vector.  

\begin{figure}[]
\begin{center}
\includegraphics[scale=0.25,angle=0]{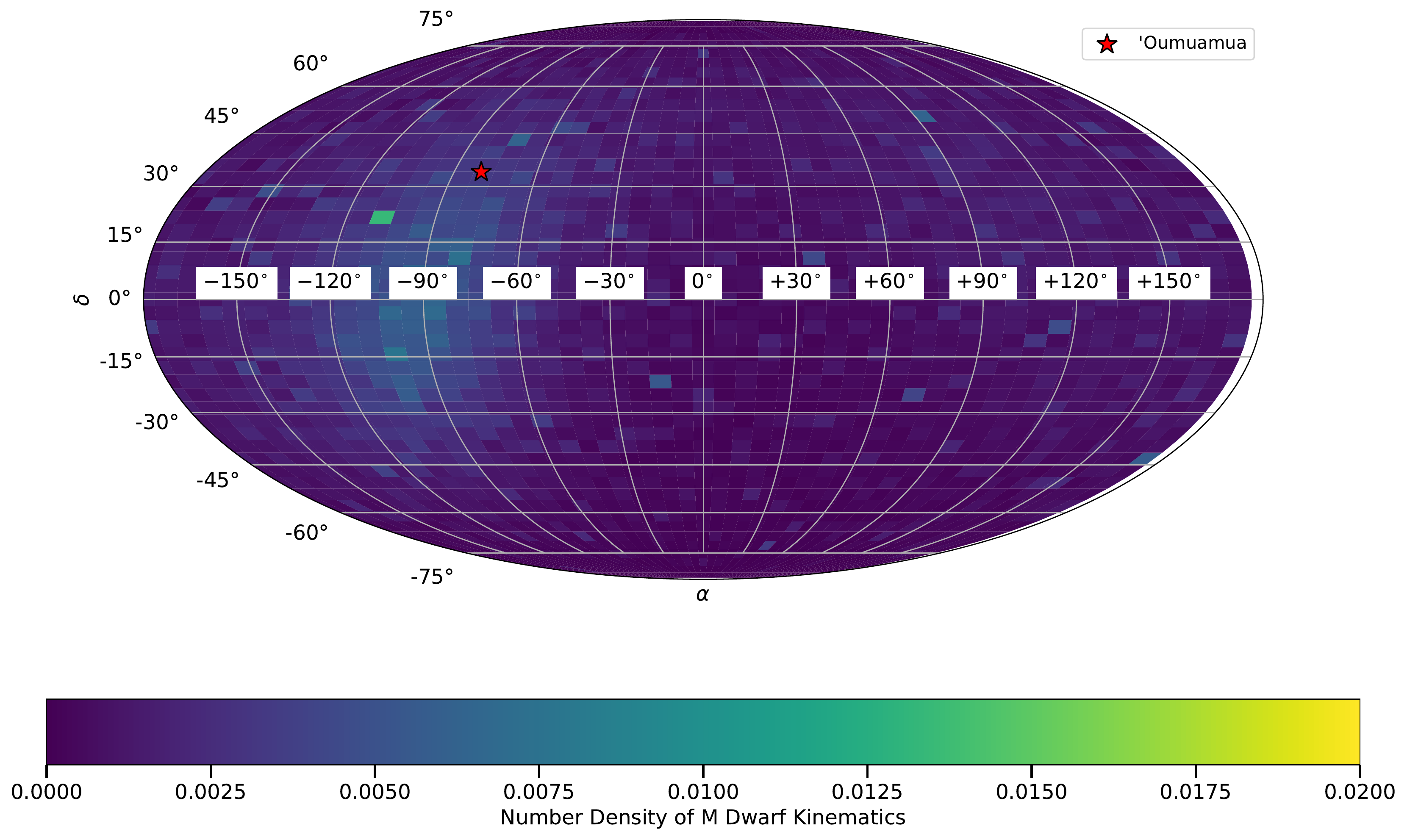}
\includegraphics[scale=0.25,angle=0]{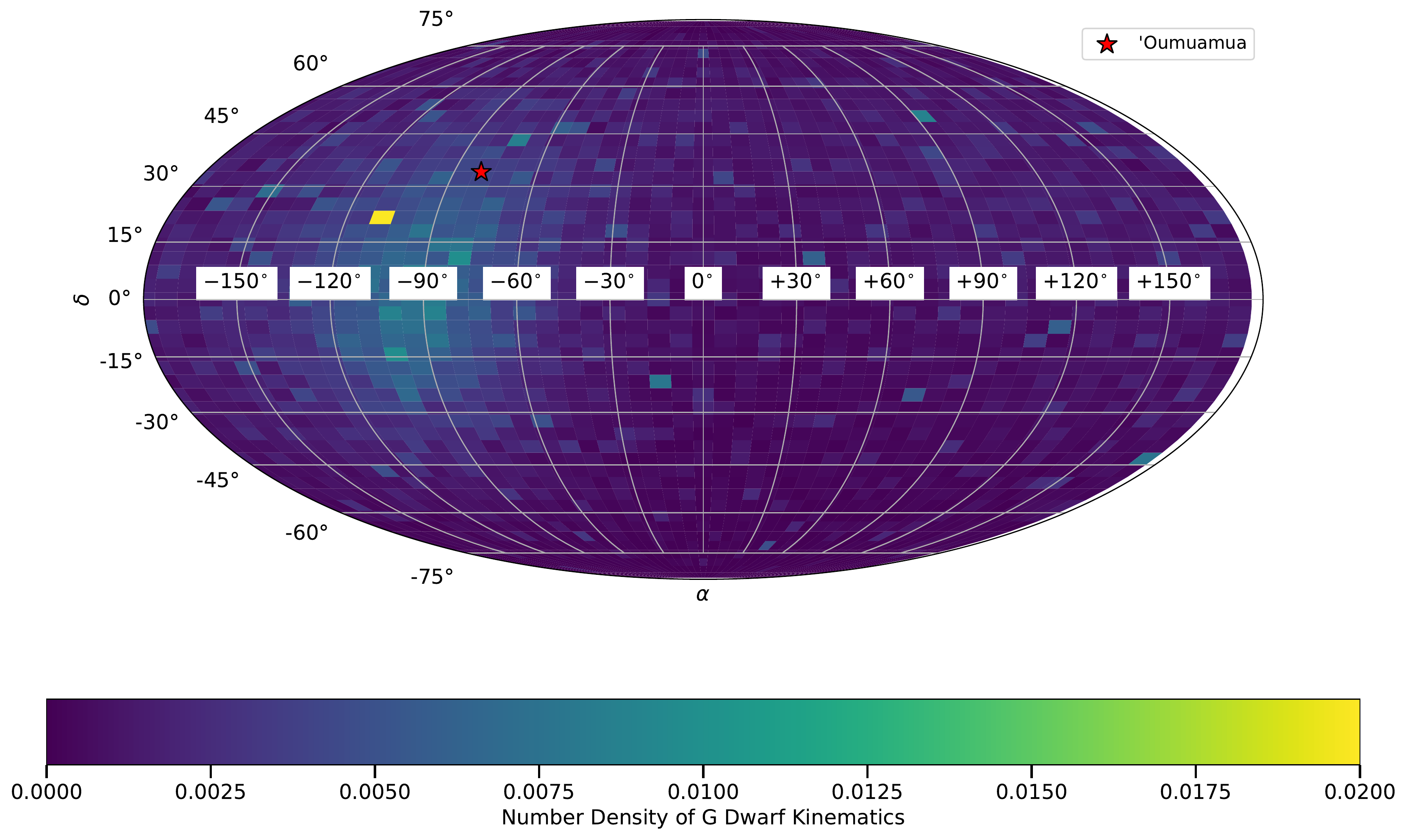}
\includegraphics[scale=0.25,angle=0]{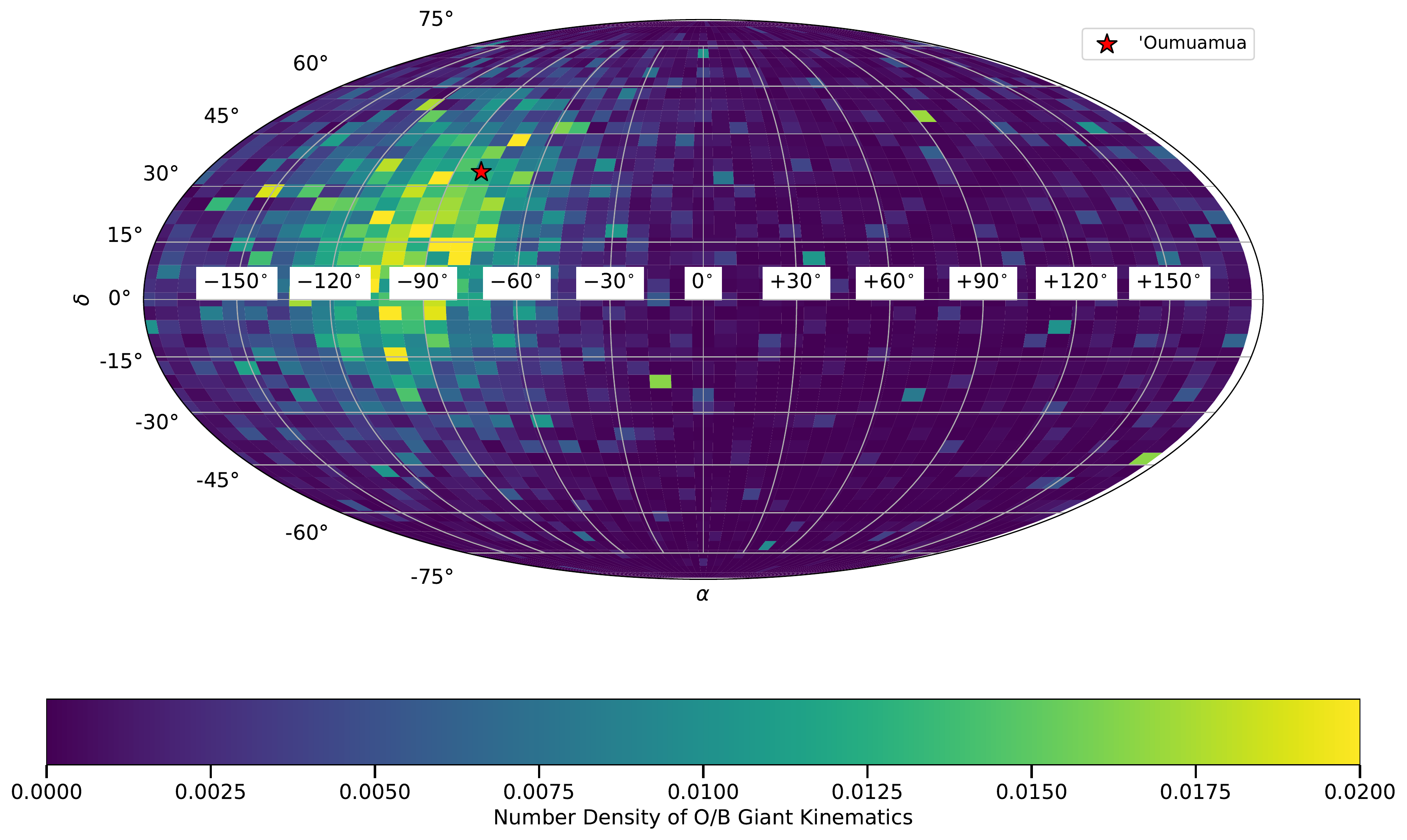}
\caption{Sky map illustrating the probability that a future interstellar object will enter the Solar System in a given direction, assuming $b=0$.  The probabilities use the Schwarzschild distributions for M dwarfs (upper panel), G dwarfs (middle panel), and O/B giants (lower panel).  Warmer colors indicate a higher probability.  In all panels, the initial position of `Oumuamua is indicated with a red star.   The majority of interstellar objects enter the Solar System from the direction of the solar apex.}
\label{iso_dist}
\end{center}
\end{figure}

An incoming interstellar object will have a velocity at $r=\infty$ drawn from this distribution, and will approach the Sun at a given impact parameter, $b$. If we assume that the impact parameter is $b=0$, then as the object enters the vicinity of the Sun, its position vector is parallel and opposite to its velocity vector.  Therefore, every velocity vector isomorphically maps to a unit vector pointing away from the Sun towards a location on the unit sphere (which we arbitrarily define as the 1 au unit sphere).  We convert each incoming heliocentric velocity to a corresponding declination and right ascension. We show the resulting sky position probability distributions  in Figure \ref{iso_dist} for kinematics representative of three different stellar populations, described in Table \ref{schwarzschild_parameters}.  The majority of interstellar objects enter the Solar System in the direction of the solar apex, regardless of the assumed stellar population.  This is because the apex is the direction with which the Sun is moving with respect to the LSR.


\begin{table}[]
\begin{tabular}{||c c c c c||}
\hline
Stellar Type & $\sigma_R$ & $\sigma_\phi$  & $\sigma_Z$  & $v_d (^\circ)$\\
\hline\hline
M & 31 & 23 & 16 & 7\\
\hline
G & 26 & 18 & 15 & 12\\
\hline
O/B & 12 & 11 & 9 & 36\\
\hline
\end{tabular}
\caption{Parameters in the Schwarzschild distributions for different stellar types.  All velocities are expressed in kilometers per second.  Since different stellar populations correspond to different velocity distributions, the initial locations of our simulated objects depend on the distribution from which we draw samples.}  
\label{schwarzschild_parameters}
\end{table}

To generate samples from our grid, we define a multivariate spline that maps velocity vectors to their associated probabilities.  For each sample, we draw a probability from a uniform distribution with bounds marked by the minimum and maximum values of probabilities associated with each point on the grid.  Then, we use differential evolution to find the velocity vector that minimizes the difference between the spline interpolation and probability.  We repeat this process $10^6$ times, generating a sample of 1,000,000 objects.  We verified that the distribution of our simulated objects was converged for this sample size.   

\begin{figure}
    \centering
    \includegraphics[scale=0.4,angle=0]{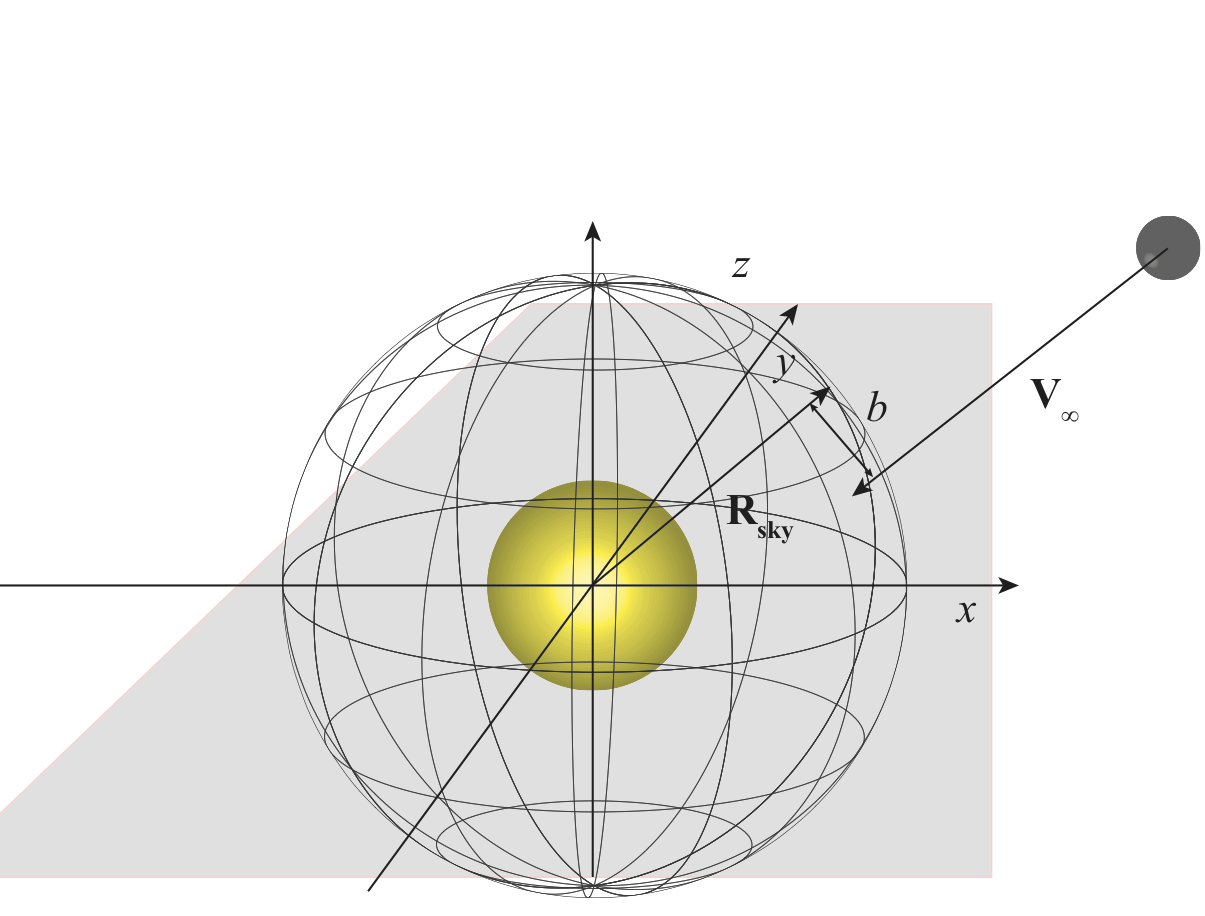}
    \caption{A schematic diagram showing the geometry of an incoming interstellar object. Each object has an initial velocity vector drawn from the galactic distribution, which maps to a location on the unit sphere in the heliocentric reference frame. We then place the interstellar object at a position given by an impact parameter, $b$, that we draw.}
    \label{Fig:schematic}
\end{figure}

Using the distribution of incoming velocity trajectories, we construct a realistic distribution of incoming interstellar object trajectories.  We denote our initial velocity vector as $\vec{v}_{\infty}$, where the magnitude, $|\vec{v}_{\infty}|$, is the hyperbolic excess velocity.  For the simplest case where $b=0$, $\vec{v}_{\infty}$ is oriented in the direction of the Sun. The incoming position vector, $\vec{R}$, is given by,

\begin{equation}\label{R_no_b}
    \vec{R} = -k\vec{v}_{\infty}\, ,
\end{equation}
for some constant, $k\in\mathbb{R}$.  In reality, each incoming interstellar object will have a nonzero impact parameter.  For every initial velocity vector in our synthetic population, we draw $b$ from a parabolic probability distribution, $p(b)\sim b^{2}$ with lower and upper bounds at 0 and 5 au, respectively. The parabolic distribution is a product of the cylindrical radial  cross section of the Solar System.  We also draw  an angle, $\psi$, from a uniform distribution from $(0, 2\pi)$ to represent the direction of $b$ in the plane normal to the velocity vector. In other words, $\psi$ represents a rotation within the tangent bundle of the 5 au sphere. We require that each interstellar object be placed at an initial position exactly $R=5$ au away from the Sun.  Since we assume that the synthetic population consists of `Oumuamua-like objects without bright cometary activity, it is unlikely that they will be detected by the LSST at large distances.  For the purposes of this paper, it is unnecessary to simulate any objects outside 5au.  Although outside the scope of this paper, it would be straightforward and worthwhile to create a synthetic population of interstellar comets that included the effects of cometary brightening, similar to the analysis presented in \citet{Cook2016}.  In Figure \ref{Fig:schematic}, we show a schematic diagram of the geometry.   We can substitute $\vec{R}_{sky}$ for $\vec{R}$ in Equation \ref{R_no_b} and solve for $k$ using,

\begin{equation}
    \vec{R}_{sky} = -\,\bigg(\sqrt{R^{2} - b^{2}}\bigg)\,\,\hat{v}_{\infty}\,,
\end{equation}
where $\hat{v}_{\infty}$ is the unit vector pointing in the direction of $\vec{v}_{\infty}$.  We define $\vec{b}$ as the vector with magnitude $b$ that lies in the tangent plane rotated by $\psi$ and has the property $\vec{b}\cdot\vec{v}_{\infty}=0$.  We can write $\vec{b}$ as 

\begin{equation}
\vec{b}=
\begin{bmatrix}
b\cos{\phi}\cos{\psi}\\
b\cos{\phi}\sin{\psi}\\
b\sin{\phi}\\
\end{bmatrix}\,,
\end{equation}
where the angle $\phi$ is given by,
\begin{equation}
\tan\phi = -\,\bigg(\frac{\cos{\psi}\,v_{\infty, x} + \sin{\psi}\,v_{\infty, y}}{v_{\infty, z}}\bigg)\,.
\end{equation}
Here the subscripts on $v_{\infty}$ indicate the component of the vector. This  allows us to construct $\vec{b}$ for each initial velocity vector.  The initial position vector, $\vec{R},$ is given by

\begin{equation}
\vec{R} = \vec{R}_{sky} + \vec{b}\,.
\end{equation}

We place each interstellar object at an initial distance of 5 au away from the Sun in the direction of $\hat{R}$. In order to account for the Solar acceleration, we multiply each velocity vector by a factor $\sqrt{v_{\infty}^2 + v_{esc}^2}/v_{\infty}$, where $v_{esc}$ is the solar escape velocity at 5 au. 

\section{Numerical Simulations}\label{section:MC_integration}

For a given  interstellar object initial condition, we use the N-body code REBOUND \citep{RL2012} to integrate the trajectories of the Earth and the interstellar object for a 3-year time period. This timespan is sufficient to capture the interstellar object's trajectory through the inner solar system, which we define as the 5 au sphere centered at the Sun.  By comparison, an object on a parabolic orbit with perihelion at 1 au crosses the 5 au sphere in $\sim 2$ years.  Given the variability of the hyperbolic excess velocities within our sample, a higher integration time is needed to accommodate ISOs with low initial velocities and higher perihelia.  For computational efficiency, we neglect the contributions from the other planets in the Solar System. In reality, the giant planets will provide gravitational perturbations to a small subset of these interstellar objects. In the extreme case, \citet{Siraj2019} investigated the efficiency with which  Jupiter captures interstellar objects. However, it has been shown that there is no evidence for bound interstellar objects currently in the Solar System \citep{Morbidelli2020}. \citet{Hands2020} demonstrated that the volume capture rate of interstellar objects was 0.051 ${\rm au^3\, yr^{-1}}$, of which only $0.033\%$  are within 6 au at any time. The initial condition that we use could  be modified to investigate the possibility of capture for objects that start at further distances, as studied by \citet{Napier2021}, which is outside of the scope of this paper. In any case, the majority of the interstellar objects in our simulated population ($>99\%$) will be unaffected by perturbations from the giant planets, which have small interaction cross sections in comparison to the 5 au sphere.

An  object with absolute magnitude, $H$, has an  apparent magnitude, $m$, which may be calculated using,

\begin{equation}\label{eq:1}
m = H + 2.5\log_{10}{\bigg(\frac{d_{BS}^2d_{BO}^2}{p(\theta)d_{OS}^4}\bigg)} \,.
\end{equation}
The parameters  $d_{BO}$,  $d_{OS}$, and $d_{BS}$ represent the distances between the body and observer,  Sun and observer, and body and Sun respectively.  $\theta$, the phase angle, and $p(\theta)$, the phase integral, are defined by

\begin{equation}\label{eq:2}
    \cos{\theta} = \frac{d_{B0}^2 + d_{BS}^2 - d_{OS}^2}{2d_{BO}d_{BS}}\,,
\end{equation}
and
\begin{equation}\label{eq:3}
    p(\theta) = \frac{2}{3}((1-\frac{\theta}{\pi})\cos{\theta} + \frac{1}{\pi}\sin{\theta})\,.
\end{equation}

Since only two interstellar objects have been detected, the absolute magnitude and size-frequency distribution of the population is unconstrained. Therefore, we assume that all interstellar objects in our sample have the same value of $H$ as `Oumuamua, $H=22.4$. In reality, the absolute magnitude will depend on the size of the body. Moreover, if an interstellar object exhibits cometary activity like Borisov, then it will be much brighter and the absolute magnitude should change as a function of heliocentric distances. For the purposes of this paper, we conservatively evaluate the detection capabilities for the LSST for `Oumuamua-like objects. However, these distributions can be  updated to account for differences in intrinsic brightness, once the size-frequency distribution of the population is better constrained.

We initialize the position of the Earth in our simulations from a distribution of 24  start dates evenly distributed between January 1st and December 31st of 2021, although the year is arbitrary.  For each simulated trajectory, we randomly select one of these dates, which defines the initial condition of the Earth.  At 100 evenly-spaced times within the 3 year integration period, we evaluate the position, velocity, apparent magnitude as seen from Earth, declination/right ascension, altitude/azimuth, orbital elements, as well as other parameters, for each interstellar object. Note that we implemented a cubic interpolation to reconstruct aspects of the trajectories for some of the results presented in \S \ref{section:Detectability}. We perform these calculations for a synthetic population of $\sim10^6$ interstellar objects. We verified that the shape of the distributions presented in the following two sections and fractional percentages presented in Table \ref{percent_table} were numerically converged.

\begin{figure*}
    \centering
    \includegraphics[scale=0.5,angle=0]{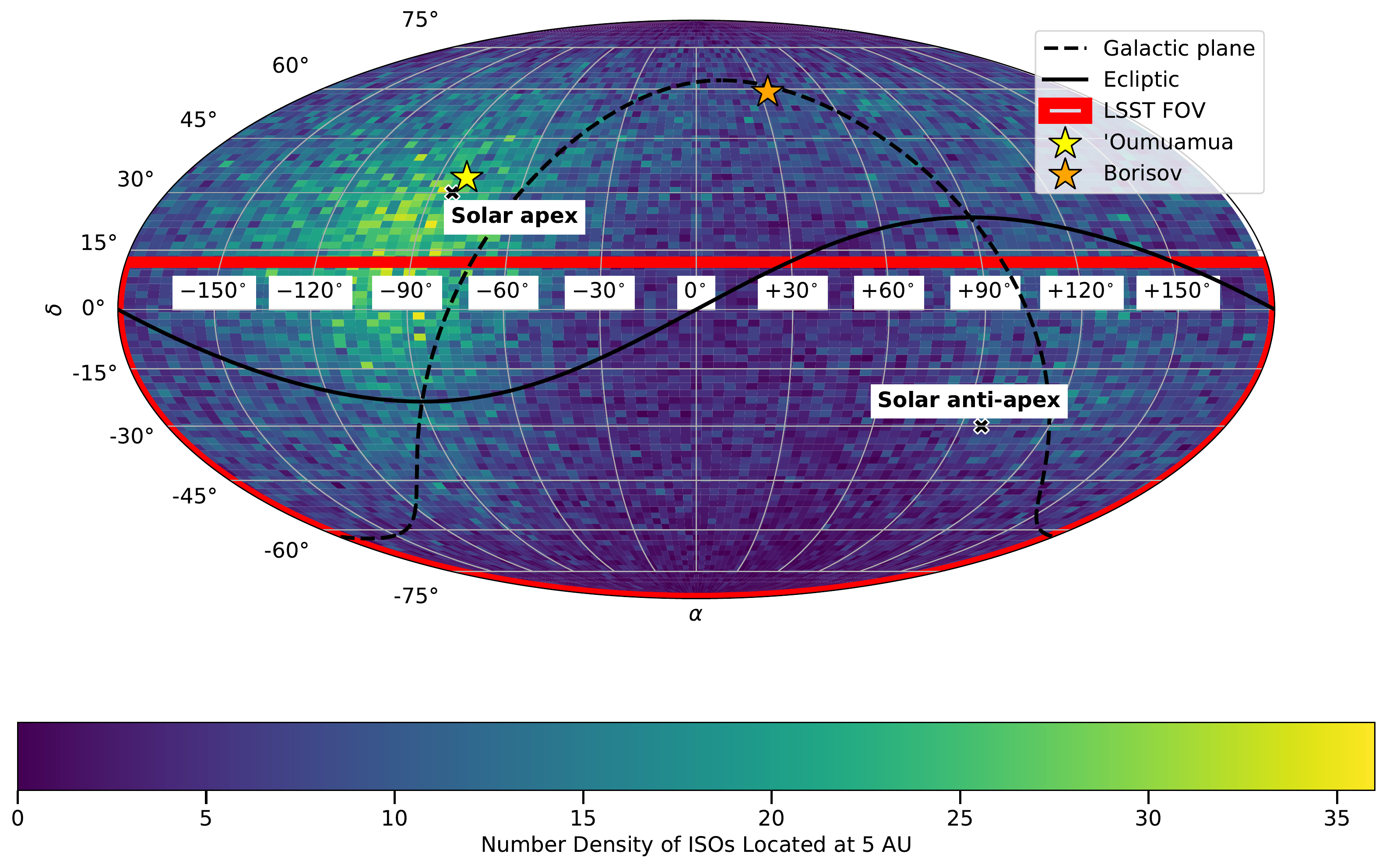}
    \caption{The initial positions of interstellar objects that are detectable by the LSST (i.e. their positions when they enter the 5 au sphere prior to detection).  We list our detectability criteria in Section \ref{section:Detectability}.  While the majority of simulated interstellar objects enter the inner Solar System in the direction of the solar apex, a significant yet  smaller fraction enter from the  anti-apex.  We include the ecliptic, the LSST's nominal field-of-view, and the positions of `Oumuamua and Borisov at the moment that they entered the Solar System for reference.  While the LSST's observational strategy is complex, with the Northern Ecliptic Spur extending to +$30^\circ$ declination only for some right ascensions \citep{Bianco2021}, the Wide-Fast-Deep, which includes all right ascensions, extends to +$12^\circ$ in most regions.  Thus, we adopt +$12^\circ$ as the LSST's nominal field-of-view.}
    \label{Fig:origin}
\end{figure*}

\begin{figure*}
    \centering
    \includegraphics[scale=0.5,angle=0]{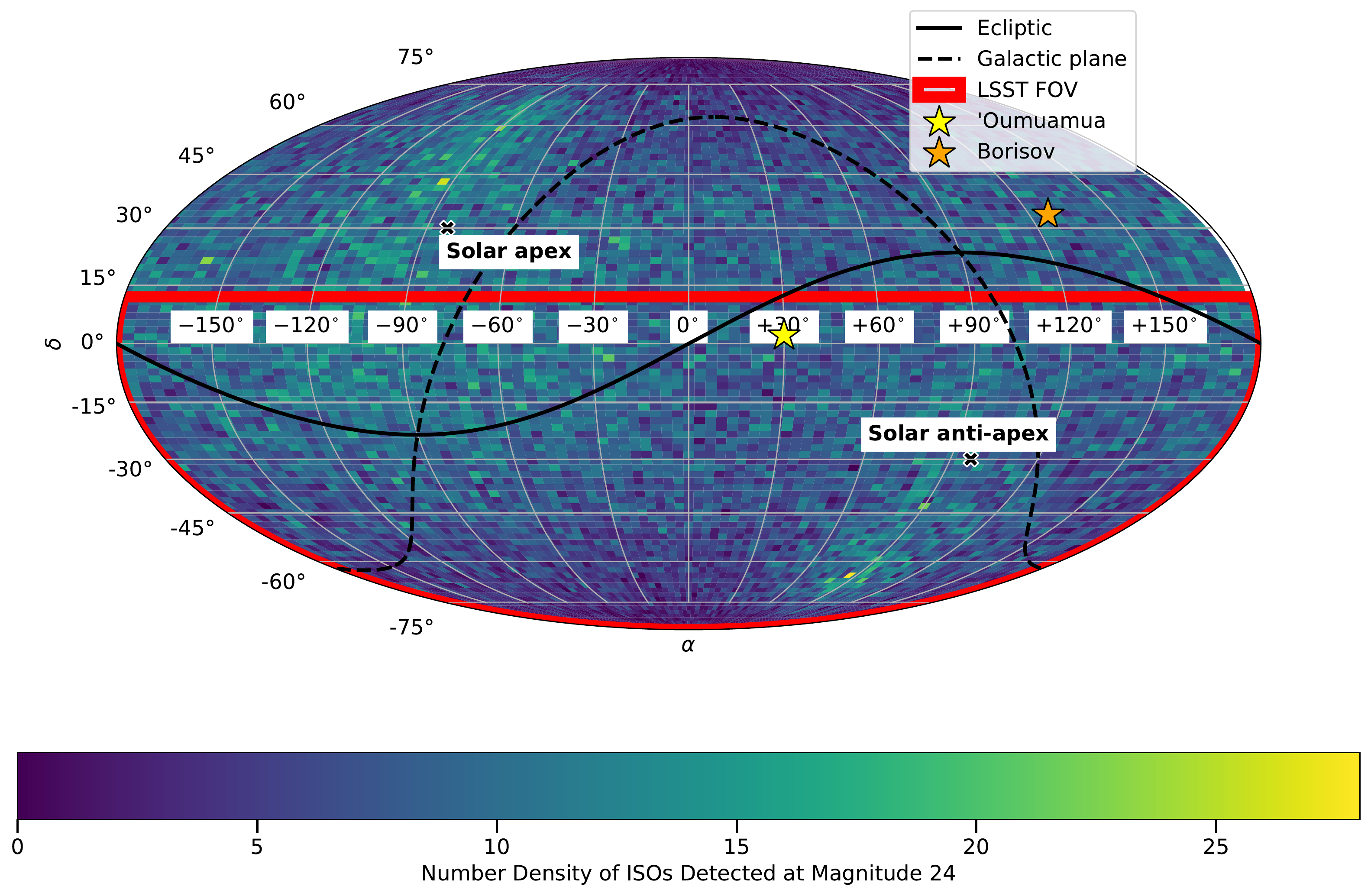}
    \includegraphics[scale=0.5,angle=0]{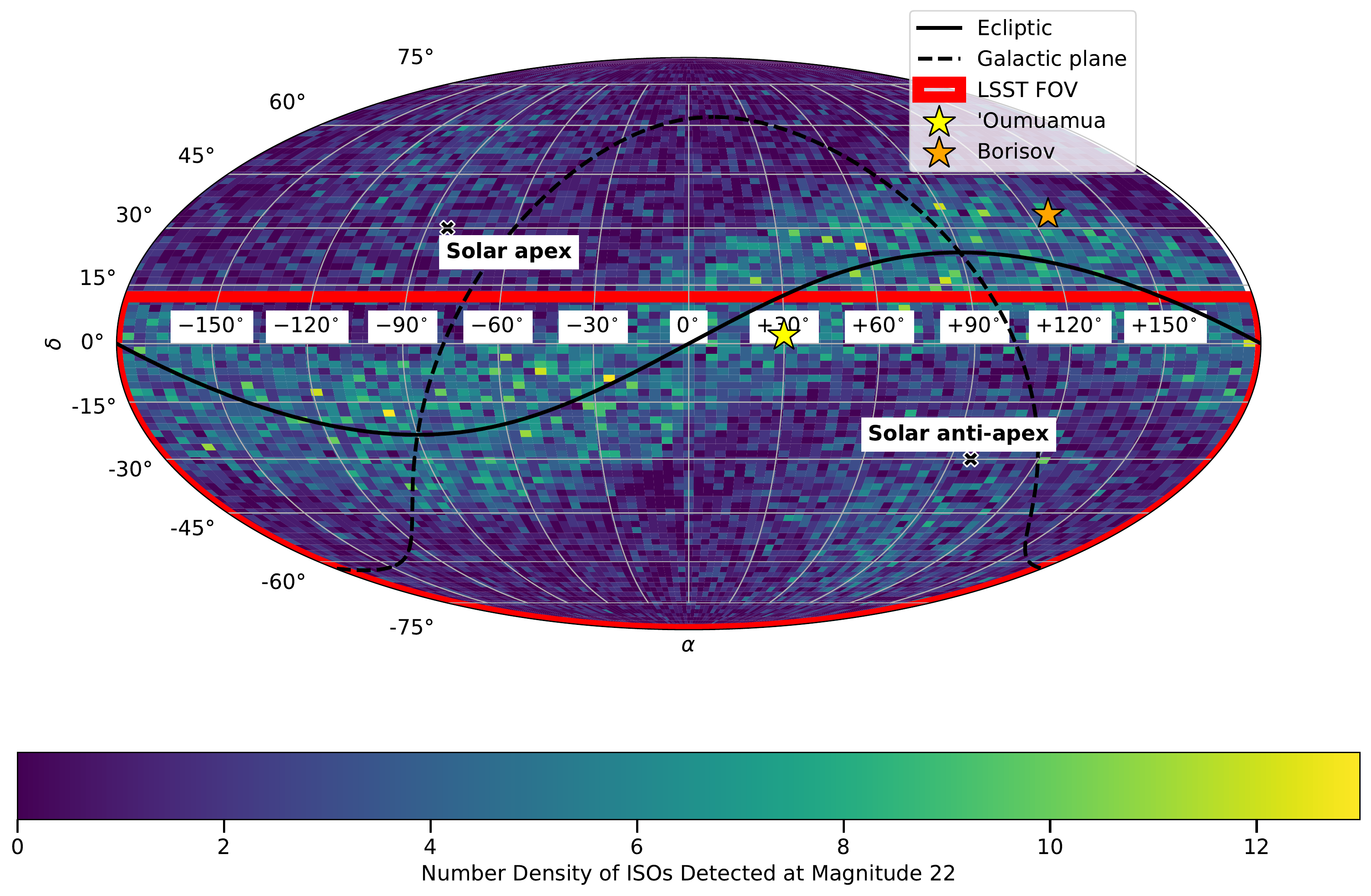}
    \caption{Sky positions of interstellar objects when they are detectable for all sky surveys with limiting magnitudes of 24 and 22. 
    The top panel shows histograms of the sky positions for the simulated objects that satisfy our detectability criteria for the LSST the moment that they reach $m\leq 24$. The nominal field of view of the LSST is over-plotted in red lines.  Compared to Figure \ref{Fig:origin}, the detectable objects  are less clustered at the  apex due to their subsequent motion along their trajectories prior to detection. 
    The  bottom panel displays the positions of objects that reach  $m\leq 22$ and  satisfy our detectability criteria, when they reach $m\sim 22$.   The brightest objects ($m\leq 22$) are preferentially found along/near the ecliptic. The positions of `Oumuamua and Borisov when they were detected are shown, which is distinct from their positions in Figure \ref{Fig:origin}.
    }
    \label{detectable}
\end{figure*}

\section{The Population Detectable with the LSST}\label{section:Detectability}

In this section,  we identify which members of the simulation  will be detectable by the LSST, given its detailed observational capabilities.    We accomplish this  using the following procedure.  For each  simulated object, we calculate the altitude/azimuth of the object and the Sun at each point along its 3-year trajectory.  We consider an interstellar object  \textit{detectable} by the LSST if it meets all of the following 3 criteria at \textit{any} time along its trajectory through the 5 au sphere:

\begin{enumerate}
    \item Its apparent magnitude is less than 24 and higher than 16.  We consider $m=24$ to be the apparent magnitude of the dimmest objects that the LSST is capable of observing.  Because of the LSST's 30-second exposures, sources brighter than the 16th magnitude would saturate, meaning they would not be detected by the LSST.
    \item Its altitude is higher than $30^\circ$ relative to the latitude/longitude of Rubin Observatory, and its declination is less than $+12^\circ$.  This ensures that the object is within the nominal field of view of the LSST, as the Wide-Fast-Deep (WFD) extends to $+12^\circ$ declination for most right ascensions, according to Figure 1 of \citet{Bianco2021}.
    \item The Sun's altitude is lower than $-18^\circ$ relative to the latitude/longitude of the LSST. This ensures that the object is observed before/after astronomical twilight, allowing for detailed observations of faint point sources.  
\end{enumerate}

These criteria ensure that when each object reaches $m\leq 24$, the angular distance between the object and the Sun is large.  We find that out of the $\sim10^6$ simulated interstellar objects, $6.956\%$ of them reach a minimum apparent magnitude brighter than 24, and $2.828\%$ of them are detectable by the LSST.  We  define a ``reachable'' object as one that can be intercepted with a rocket sent from Earth with an impulsive change of velocity,  $\Delta$v$<20$ km/s.  We further explore the reachable population in Section \ref{section:Intercept}.   

In Figure \ref{Fig:origin}, we show the position on the sky of the interstellar objects that are detectable with the LSST when they are \textit{entering} the Solar System, while in the top panel of Figure \ref{detectable}, we show the sky positions of the same population of objects when they reach $m\leq 24$, which corresponds to the LSST's limiting magnitude.  We include the ecliptic and galactic planes for reference, since precise observations near the latter are difficult.  It is clear that there is a strong clustering in the sky location of these objects in the vicinity of the solar apex as they enter the 5 au sphere.  At the point within the 5 au sphere that the objects first become detectable, there is still a clustering of objects towards the solar apex and anti-apex, but the detection locations are more isotropic than the locations at which they enter the 5 au sphere.  This is because the object has moved significantly through its trajectory by the time it is detectable.

In the bottom panel of Figure \ref{detectable}, we show the subset of this population that reach an apparent magnitude of $m\leq 22$. This subset would be representative of the detected population in a less sensitive observational search. There is a strong clustering of detections in the vicinity of the ecliptic as well as minor groupings of objects around the solar apex and anti-apex, which are remnants of the incoming distribution.  The distribution of locations in Figures \ref{Fig:origin} and \ref{detectable} reveal the likelihood of detecting interstellar objects with similar trajectories as `Oumuamua or Borisov.  Because of `Oumuamua's initial position near the solar apex, one would expect to detect more objects from the same region in the sky.  Borisov's initial position, far from the solar apex and anti-apex, was unusual.  Because of Borisov's high incoming velocity, its age was high.  According to Figure \ref{iso_dist}, older objects have more isotropic incoming trajectories, explaining Borisov's behavior.  However, its relatively close approach to Earth and, by extension, the ecliptic, rendered it detectable.

\begin{figure}
    \centering
    \includegraphics[scale=0.5,angle=0]{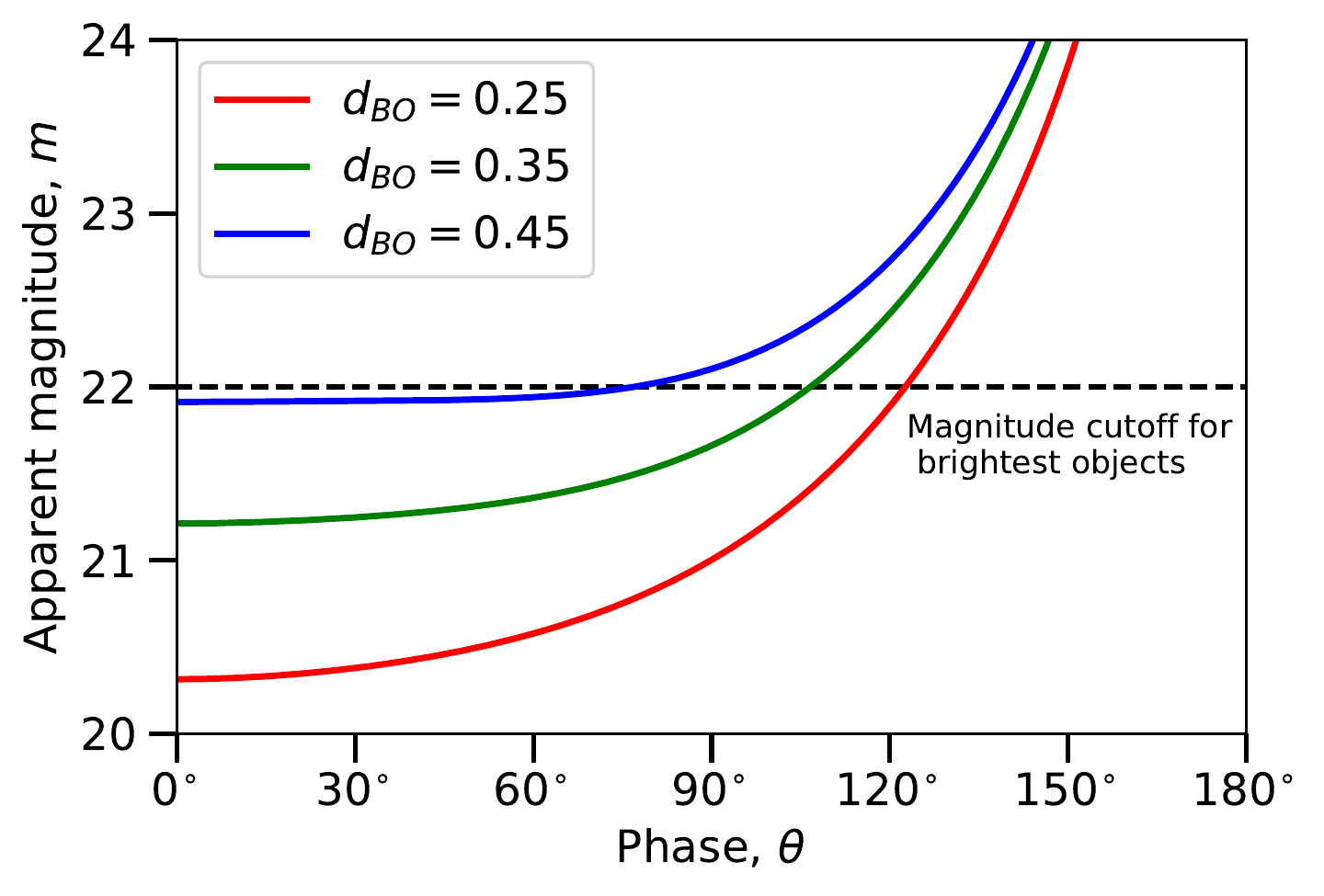}
    \caption{Apparent magnitude, $m$, as a function of phase angle, $\theta$, for an `Oumuamua like object with $d_{BO}=0.25$au (red), $d_{BO}=0.35$au (green), and $d_{BO}=0.45$au (blue).  The apparent magnitude is minimized for $\theta\approx 0^\circ$, when the object is in the ecliptic plane.  The closer an object comes to the Earth in this range, the more sensitively its brightness depends on phase angle.}
    \label{fig:ecliptic_proof}
\end{figure}

To explain the clustering of bright objects near the ecliptic, we evaluate the phase angle, $\theta$, that minimizes the apparent magnitude of an  object for different values of their minimum distance $d_{BO}$.   We calculate the apparent magnitude as a function of phase angle using Equations \ref{eq:1}-\ref{eq:3}, assuming that the Earth has a circular orbit ($d_{OS}=1$au).  In Figure \ref{fig:ecliptic_proof}  we show the apparent magnitude as a function of phase angle for three values of $d_{BO}$ which are representative of the simulated objects that attain $m\sim22$. In our simulated population, the vast majority of detected  objects do not pass close to the Earth, and hence $d_{BO}\gg R_{\oplus}$ (Figure \ref{fig:min_dist}). A phase angle of $\theta=0^\circ$ minimizes the apparent magnitude for a constant value of $d_{BO}$.  This  corresponds to the configuration where the object,  Earth, and  Sun are aligned, with the Earth in the center. Therefore, an interstellar object that achieves a distance from the Earth of $ R_{\oplus}\ll d_{BO}<0.5$au is brightest in the ecliptic plane.  The clustering around the ecliptic in the bottom panel of Figure \ref{detectable}  extends to $\sim\pm20^\circ$.  For an object with $d_{BO}=0.45$au, this corresponds to a phase angle of $\sim 50^\circ$, which, according to Figure \ref{fig:ecliptic_proof}, produces an apparent magnitude fainter than 22.   Therefore, the clustering of the brightest interstellar objects around the ecliptic is due the limiting magnitude of 22. This limits the  detected objects to the ones that come within $<0.5$au of the Earth, whose brightness depends most sensitively on the phase angle. 

\begin{figure}
    \centering
    \includegraphics[scale=0.5,angle=0]{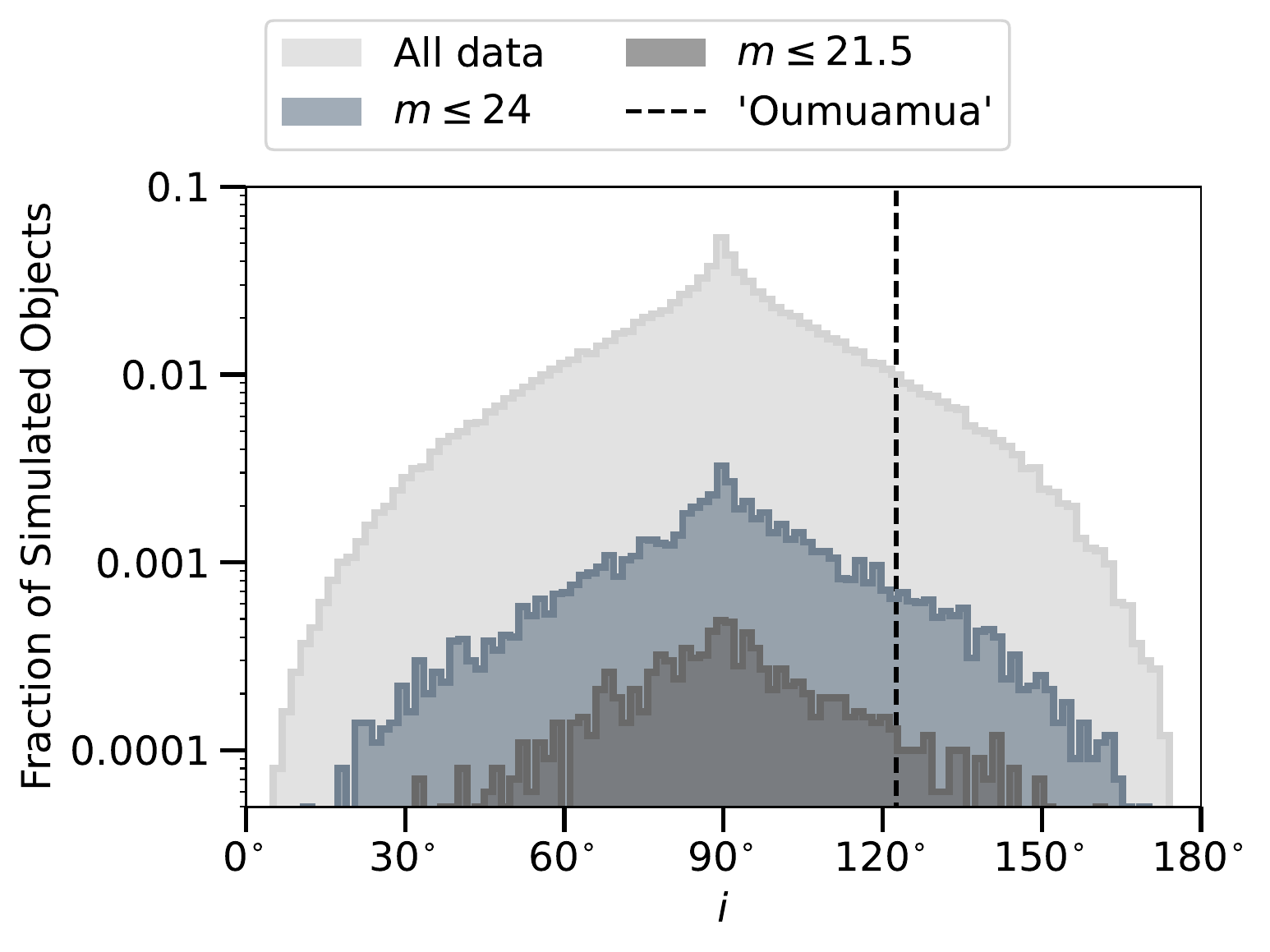}
    \caption{Orbital inclination of the initial population of interstellar objects, and all of the objects that reach apparent magnitudes of 24 and 21.5.  As the magnitude cutoff decreases, the distribution of inclinations trends towards uniformity.  Our distribution for all objects is similar to that displayed in Figure 2c of \citet{Engelhardt2014}.  A typical interstellar object enters the Solar System on a polar trajectory.}
    \label{Fig:inc}
\end{figure}

In Figure \ref{Fig:inc}, we show the distribution of orbital inclination for the entire population of interstellar objects and the objects that reach apparent magnitudes of 24, the limiting  magnitude for the LSST.  Since other surveys, such as Pan-STAARS and the Zwicky Transient Facility \citep{Chambers2016, Bellm2014}, have fainter limiting magnitudes than the LSST, we also show the distribution for objects that reach apparent magnitudes of 21.5. The initial  and detectable populations  are centered at inclinations of $\sim90^\circ$.  This  is due to the orientation and  velocity of  the Solar System with respect to the galactic mid-plane.  There are  two  compounding properties that produce inclined orbits for interstellar objects.  The first is that the ecliptic plane is inclined with respect to the galactic mid-plane by $\sim60^\circ$, and second is that  the apex is at $\sim30^\circ$ in declination. Since the shape of the inclination distributions of the initial conditions and $m\leq24,21.5$ populations are similar, we conclude that the detectability criteria is independent of the inclination.   \citet{Engelhardt2014}  found a $\sin{i}$ inclination distribution that was also maximized at $i=90^\circ$. The differences in the shape of the distributions presented here  can be attributed to the  methodology for initializing trajectories.  

\begin{figure}
    \centering
    \includegraphics[scale=0.5,angle=0]{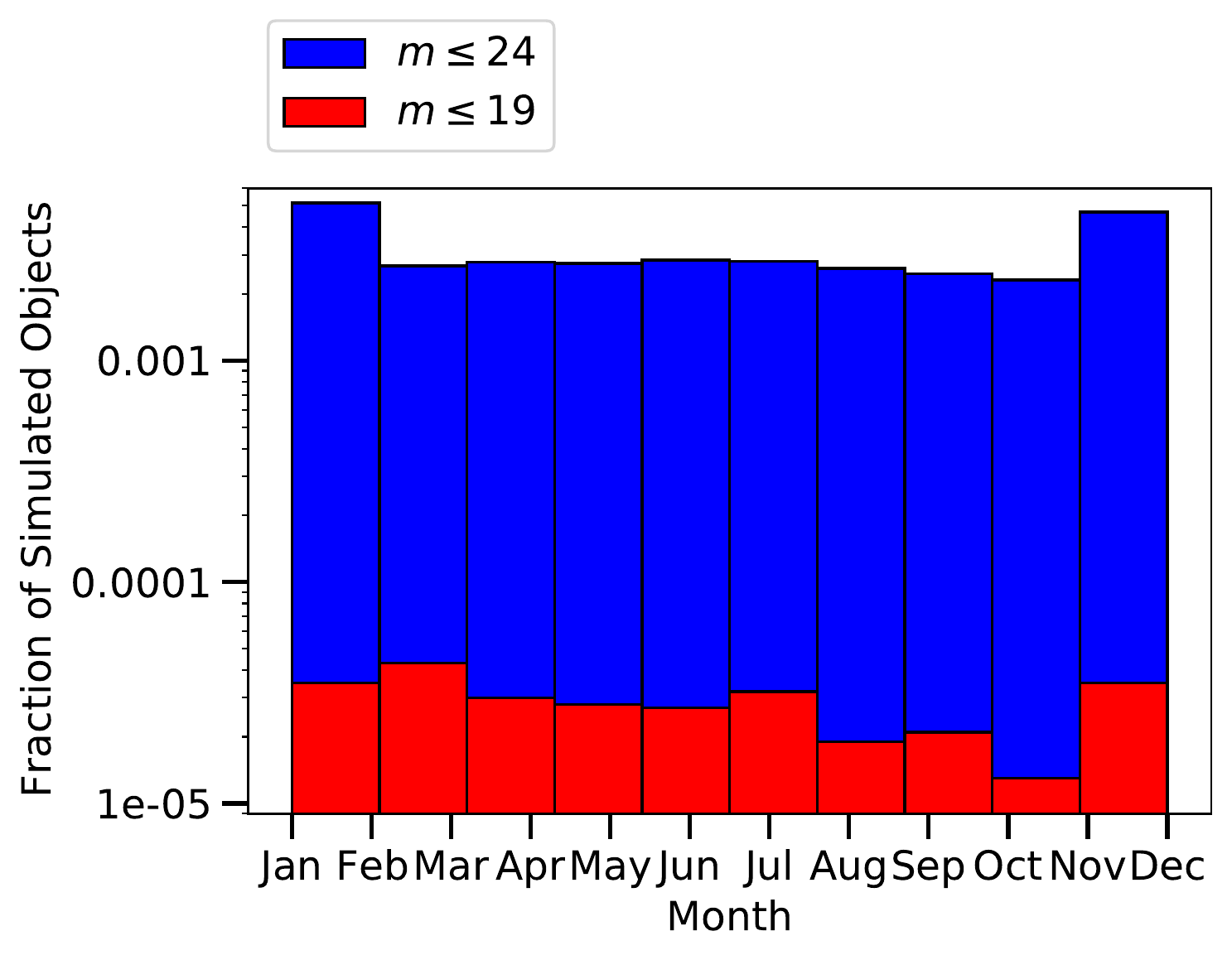}
    \caption{The seasonal variability with which interstellar objects are detected by the LSST (representing $m\leq 24$) and Pan-STARRs (representing $m\leq 19$). We show the fraction of objects that meet observability criteria and reach $m\leq24$ and $m\leq19$ as a function of month detected in blue and red histograms.   The frequency of ISO detections with the LSST is slightly higher than average close to December/January.}
    \label{fig:month_detect}
\end{figure}

\begin{figure}
    \centering
    \includegraphics[scale=0.5,angle=0]{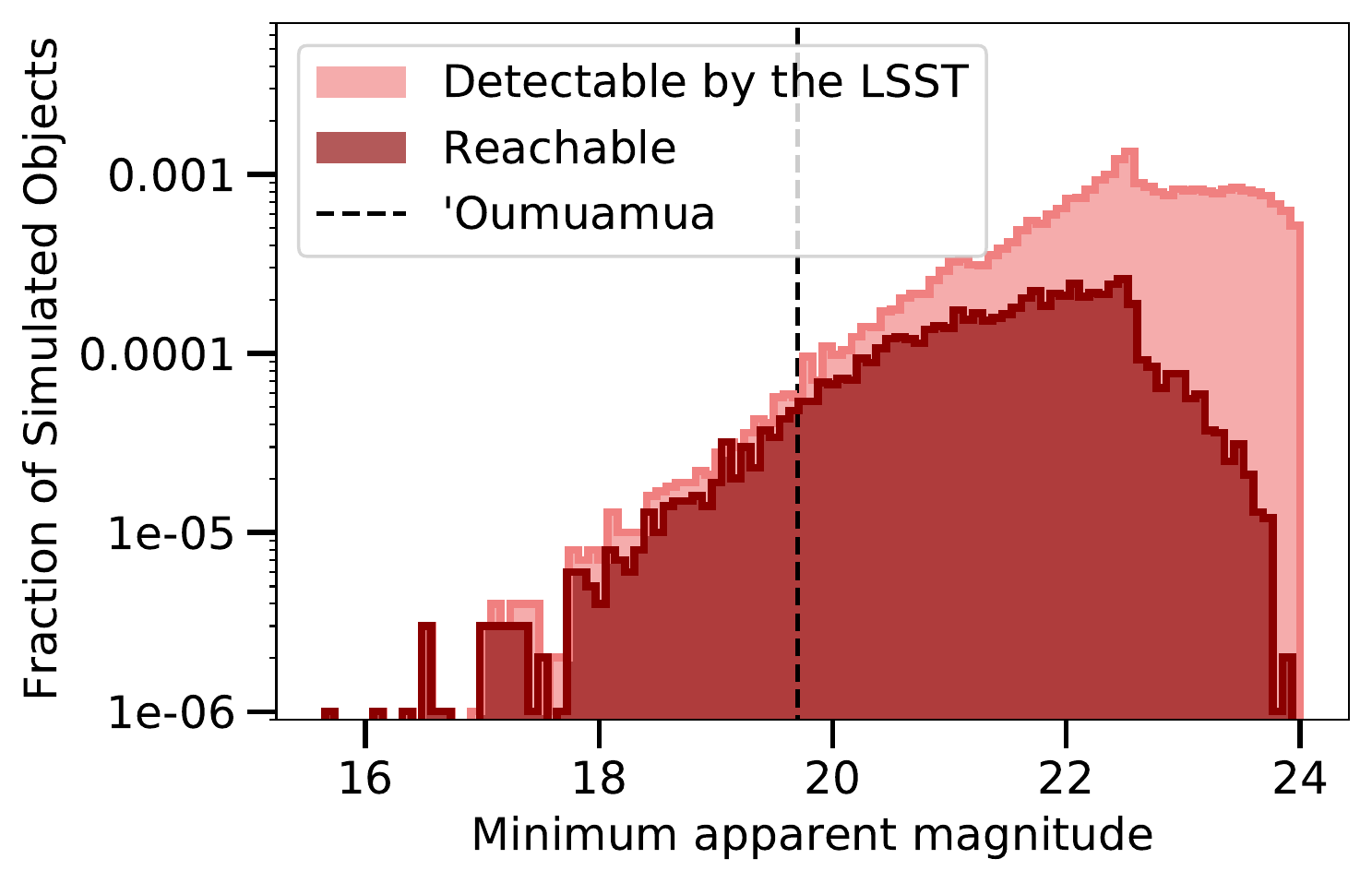}
    \caption{Minimum apparent magnitudes of each interstellar object belonging to the detectable and reachable ($\Delta v\leq 20$ km/s) populations.  The black line indicates the minimum apparent magnitude that `Oumuamua reached.  The majority of detectable ISOs do not get brighter than $m\sim19$.}
    \label{mag_histogram}
\end{figure}

\begin{figure}
    \centering
    \includegraphics[scale=0.5,angle=0]{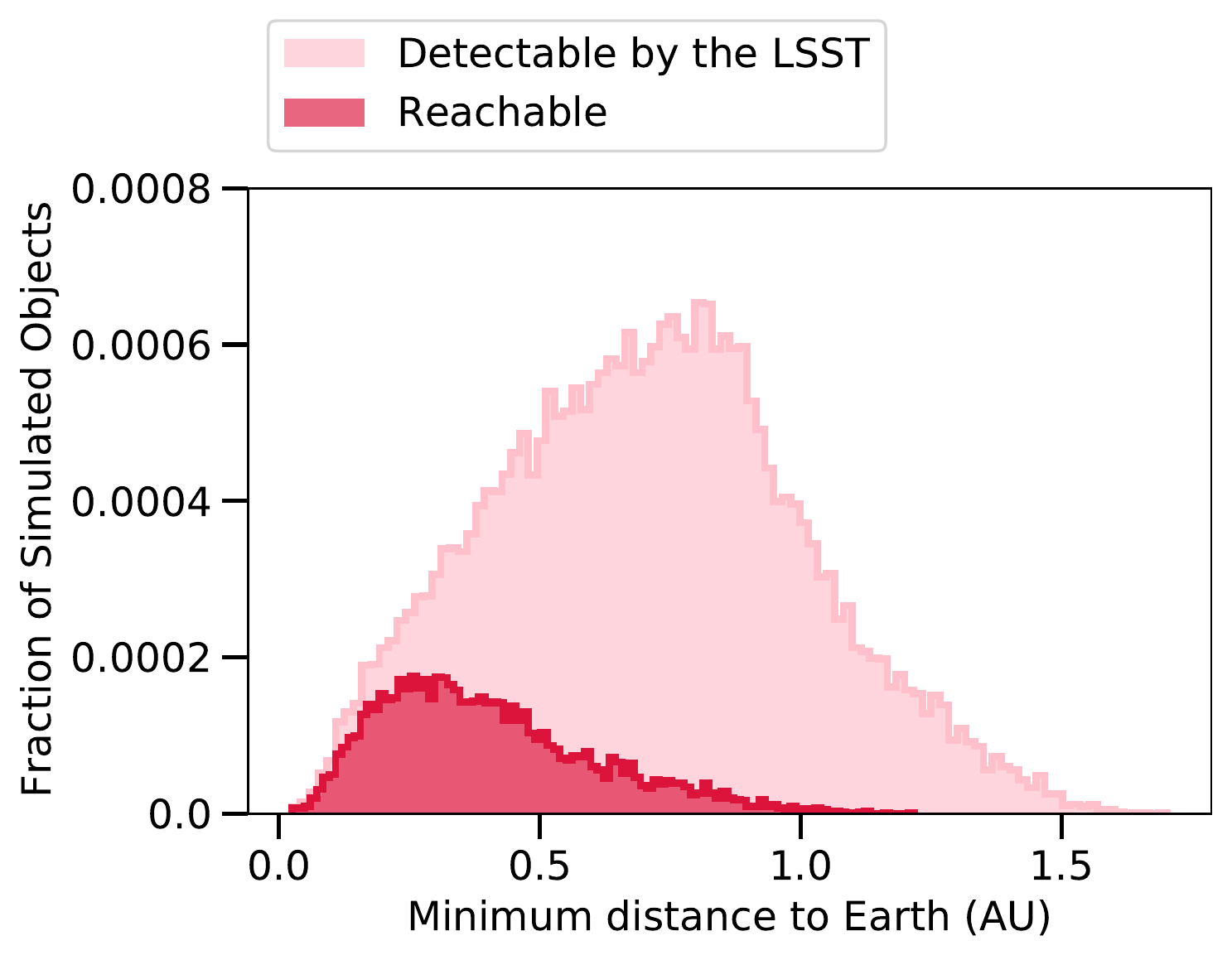}
    \caption{The minimum distances to the Earth of each interstellar object  in the detectable and reachable populations.  Detectable and reachable objects pass within $\sim 1.5$ and $\sim1$ au of the Earth, respectively.}
    \label{fig:min_dist}
\end{figure}

\begin{figure}
    \centering
    \includegraphics[scale=0.5,angle=0]{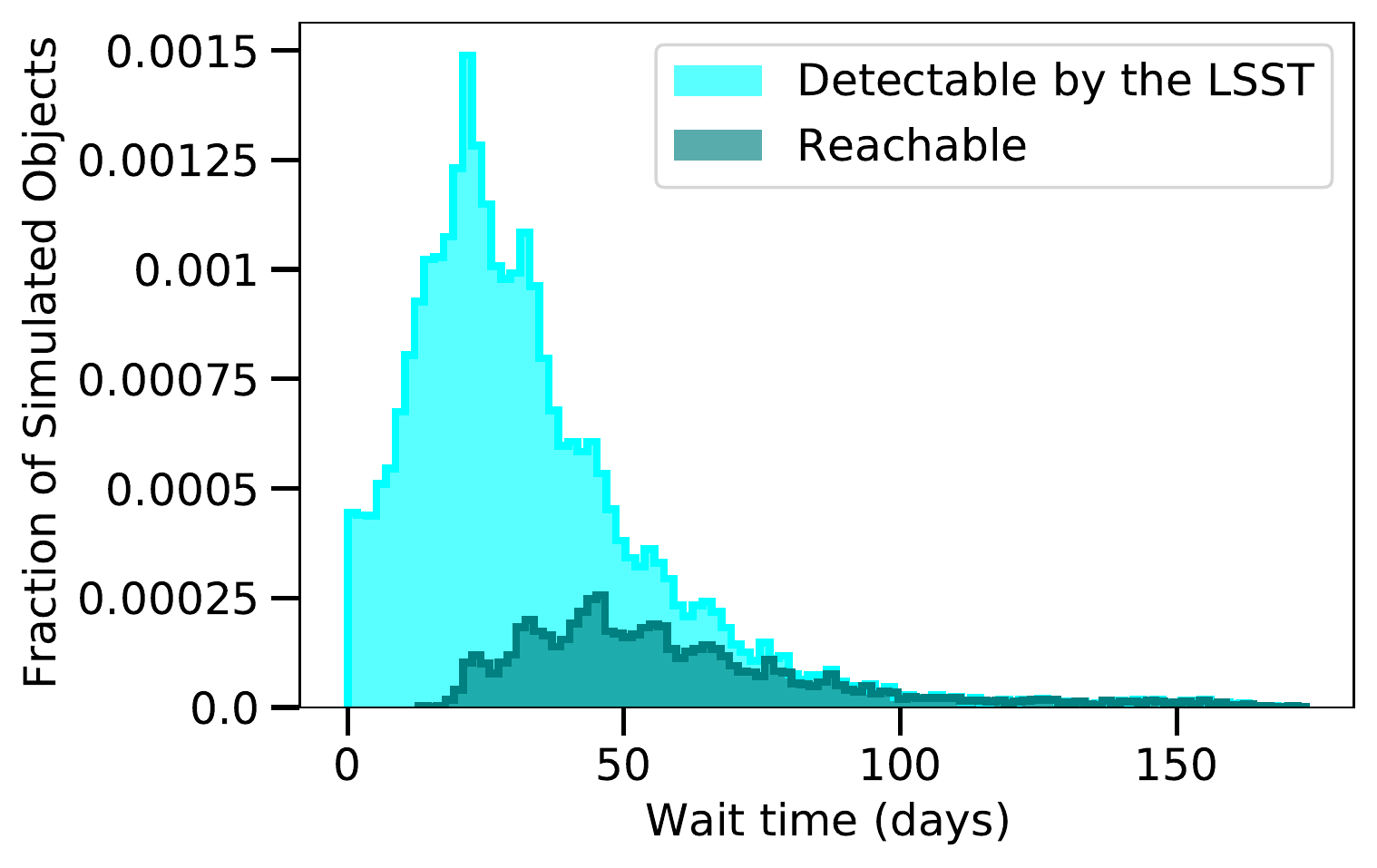}
    \caption{The wait time, or time between detection and closest approach to the Earth, of each interstellar object in the detectable and reachable populations.  Most ISOs have wait times of $\sim$50 days or less, while the upper limit  is $\sim 150$ days for both populations.}
    \label{fig:waittime}
\end{figure}

In Figure \ref{fig:month_detect}, we show the seasonal variability with which the the LSST detects objects in our simulated population. We also show the seasonal variability for the subset of these objects that reach $m<19$, corresponding to roughly the observational capabilities of currently operational all-sky surveys such as Pan-STARRs. While there appears to be a slight enhancement in the detection rate in December and January, there is no statistically significant seasonal variability in detection rate.  Our results do not account for the LSST's seasonal observational strategy.

In Figure \ref{mag_histogram}, we show histograms of the minimum apparent magnitudes of simulated interstellar objects.  Both the detectable and reachable ($\Delta v \leq 20$ km/s) populations have distributions that are skewed to fainter apparent magnitudes.  Almost all of the simulated objects ($>99\%$) within both subgroups do not get brighter than $m\sim 19$.  If `Oumuamua is a representative member of its population, then it is not surprising that we have not detected closer, brighter objects.   With the increase in sensitivity that the LSST represents, we will probe a vastly larger population of interstellar objects than ever before.

In Figure \ref{fig:min_dist}, we show the distributions of the minimum distances to the Earth for the detectable and reachable interstellar objects.  The minimum distance is calculated by extremizing a cubic spline interpolation on the relative distance as a function of time.  Detectable and reachable objects  pass within $\sim$1.5 and $\sim1$ au of the Earth, respectively.   Although the galactic population of interstellar objects should produce a $b^2$ probability distribution,  the detectable and reachable criteria preferentially  selects objects that come close to the Earth.

In Figure \ref{fig:waittime}, we show the distribution of the wait time for every object, or time between detection by the LSST and closest approach to the Earth. This distribution is roughly log normal. The LSST should detect interstellar objects before they reach their perihelia, with sufficient time to send a rendezvous mission.  It is possible that we will detect ISOs with wait times up to  $\sim$150 days.

\section{Potential Rendezvous Targets}\label{section:Intercept}

\begin{figure}
    \centering
    \includegraphics[scale=0.5,angle=0]{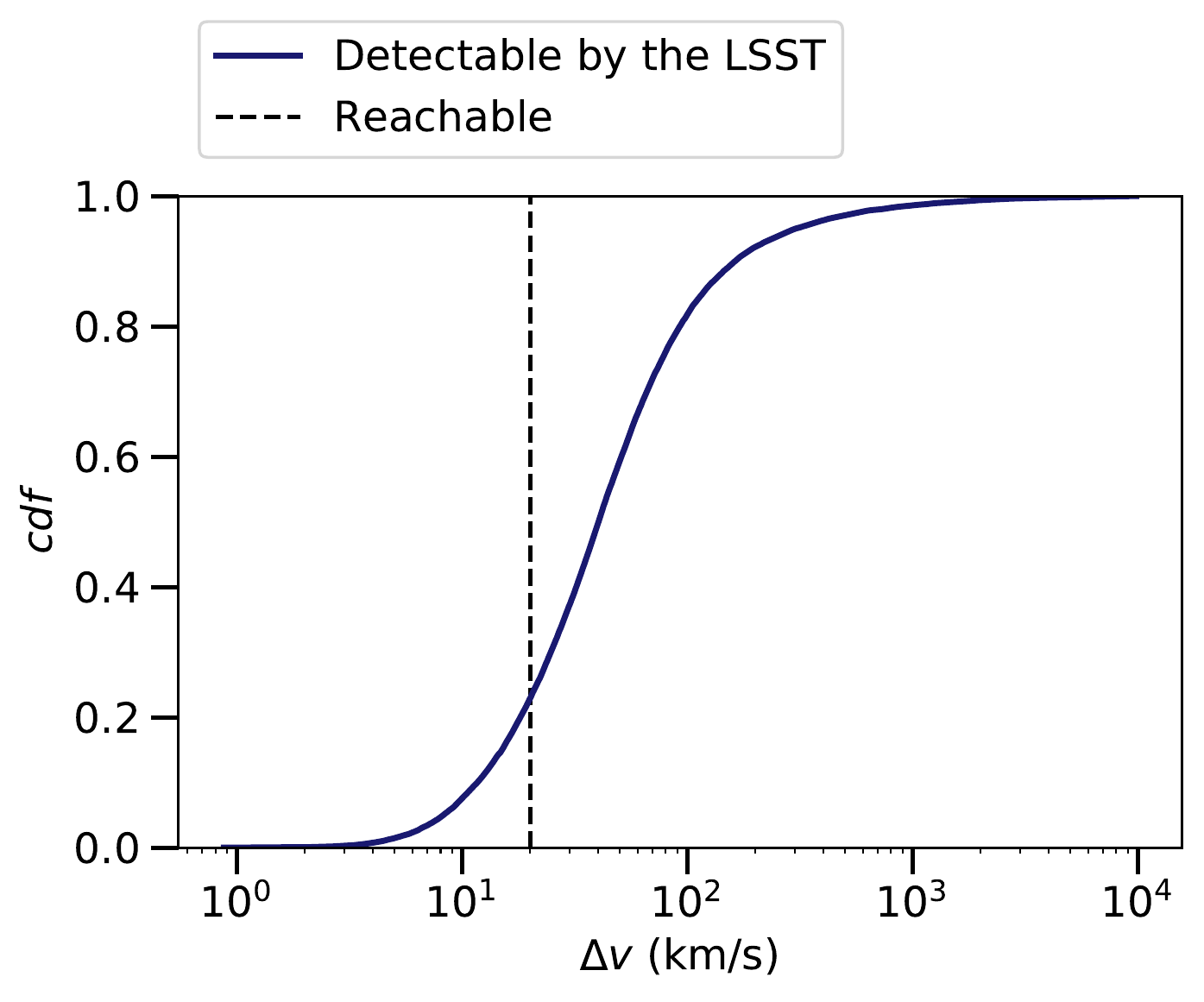}
    \caption{The $\Delta$v required to guarantee an interception for each interstellar object belonging to the detectable population.   We consider any object that satisfies $\Delta v < 20$ km/s as reachable, which corresponds to $\sim 20\%$ of detectable objects.}
    \label{fig:deltav}
\end{figure}

\begin{figure}
    \centering
    \includegraphics[scale=0.5,angle=0]{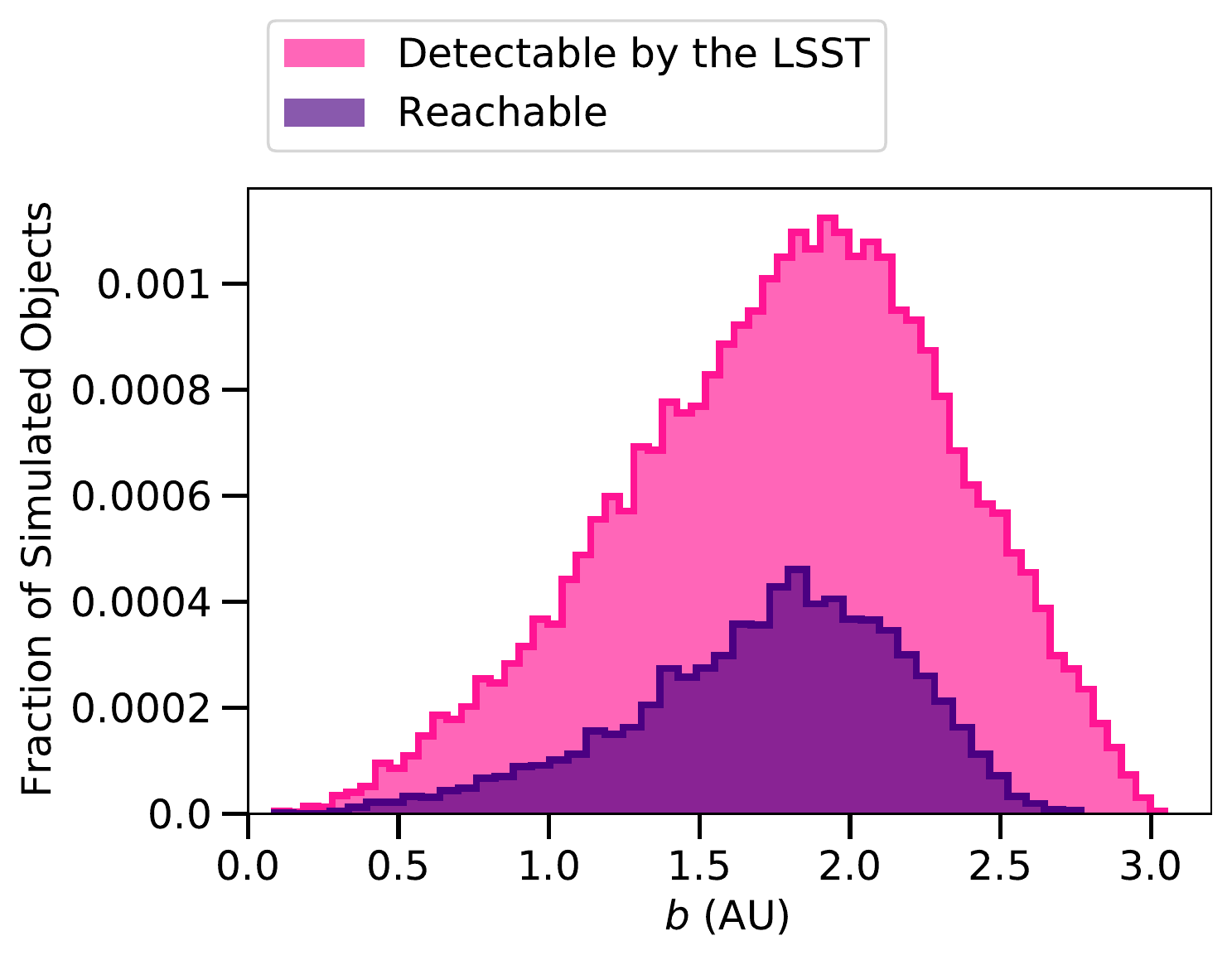}
    \caption{Distributions of dynamical impact parameters for the detectable and reachable populations.  Although the population is drawn from a $b^2$ distribution, the LSST will detect objects out to 3 au. Note that $b$ here represents the orbital impact parameter, and is not related to a hypothetical impactor mission.}
    \label{fig:b_dist}
\end{figure}

\begin{figure}
    \centering
    \includegraphics[scale=0.5,angle=0]{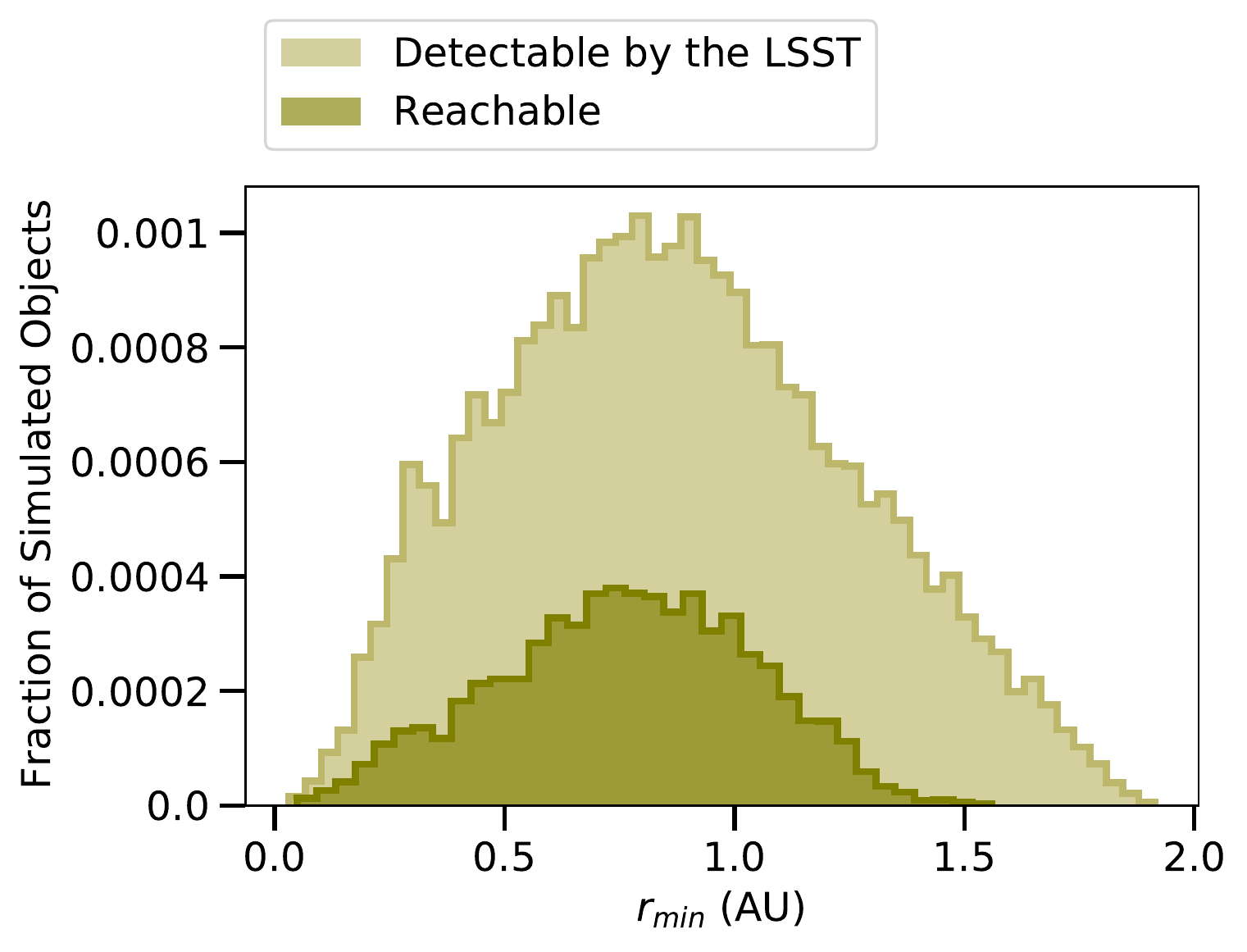}
    \caption{Distributions of perihelia for the detectable and reachable populations.  The LSST will detect  objects with perihelia out to 2 au.}
    \label{fig:rmin}
\end{figure}

In this section we further refine our simulated trajectories to examine the distribution of objects that are not only detected by the LSST but will also be feasible targets for an \textit{in-situ} interception mission.  Such a mission could include imaging, spectroscopy and, if $\Delta$v capabilities are sufficient, an impactor.  An impactor collision of this type would excavate a substantial amount of sub-surface material from an incoming interstellar object. This material could be closely examined by a companion flyby spacecraft, with a suite of instruments similar to those proposed for \textit{Bridge} \citep{Moore2021}, allowing for detailed characterization of the composition. Such a strategy was employed by the \textit{Deep Impact} mission \citep{Ahearn2005}. 

Trajectories that will be ``reachable'' from Earth can be categorized by the magnitude of the impulsive $\Delta$v required to guarantee an interception for a spacecraft sent from the Earth or the Earth's Lagrange points L1 or L2.  Such a spacecraft would execute a maneuver that would put it on an intercept trajectory before its target would reach its closest approach to Earth.  While an ISO intercept mission in this manner would be feasible only for objects with sufficiently large wait times, Figure \ref{fig:waittime} demonstrates the existence of a minimum wait time at $\sim 25$ days for the reachable population.  This would provide a mission with the time needed to design and execute a transfer maneuver. We note that this is a highly simplified estimation, and that, while outside of the scope of this paper, it would be worthwhile to investigate the  dynamics of  optimal trajectories for such a spacecraft.  \citet{Seligman2018} provided an order-of-magnitude estimate for an attainable $\Delta$v, using the quoted payload capability of a SpaceX Falcon Heavy to Mars.  They calculated that a maximum $\Delta$v$\sim 15$km/s would be attainable for such a mission concept to match the impact kinetic energy achieved by the \textit{Deep Impact} Tempel I mission.

For our simulated trajectories, we use a fiducial $\Delta v < 20$ km/s for reachable trajectories. This is meant to include the distribution of objects that may be reachable with additional $\Delta$v from deep space maneuvers. We estimate the required $\Delta$v for each of our simulated objects by dividing the minimum distance to the Earth in Figure \ref{fig:min_dist} by the wait time in Figure \ref{fig:waittime}. The resulting distribution of $\Delta$v is shown in Figure \ref{fig:deltav}.  In Figure \ref{mag_histogram}, we show the minimum apparent magnitude for the subset of objects that reach $\Delta v < 20$ km/s. We find that   $\sim 20\%$ of detectable objects are reachable for such a mission. This implies that roughly 1 in 5 interstellar objects detected by the LSST will be viable rendezvous targets. This is significantly better than the estimates in \citet{Seligman2018}.

\begin{figure}
    \centering
    \includegraphics[scale=0.5,angle=0]{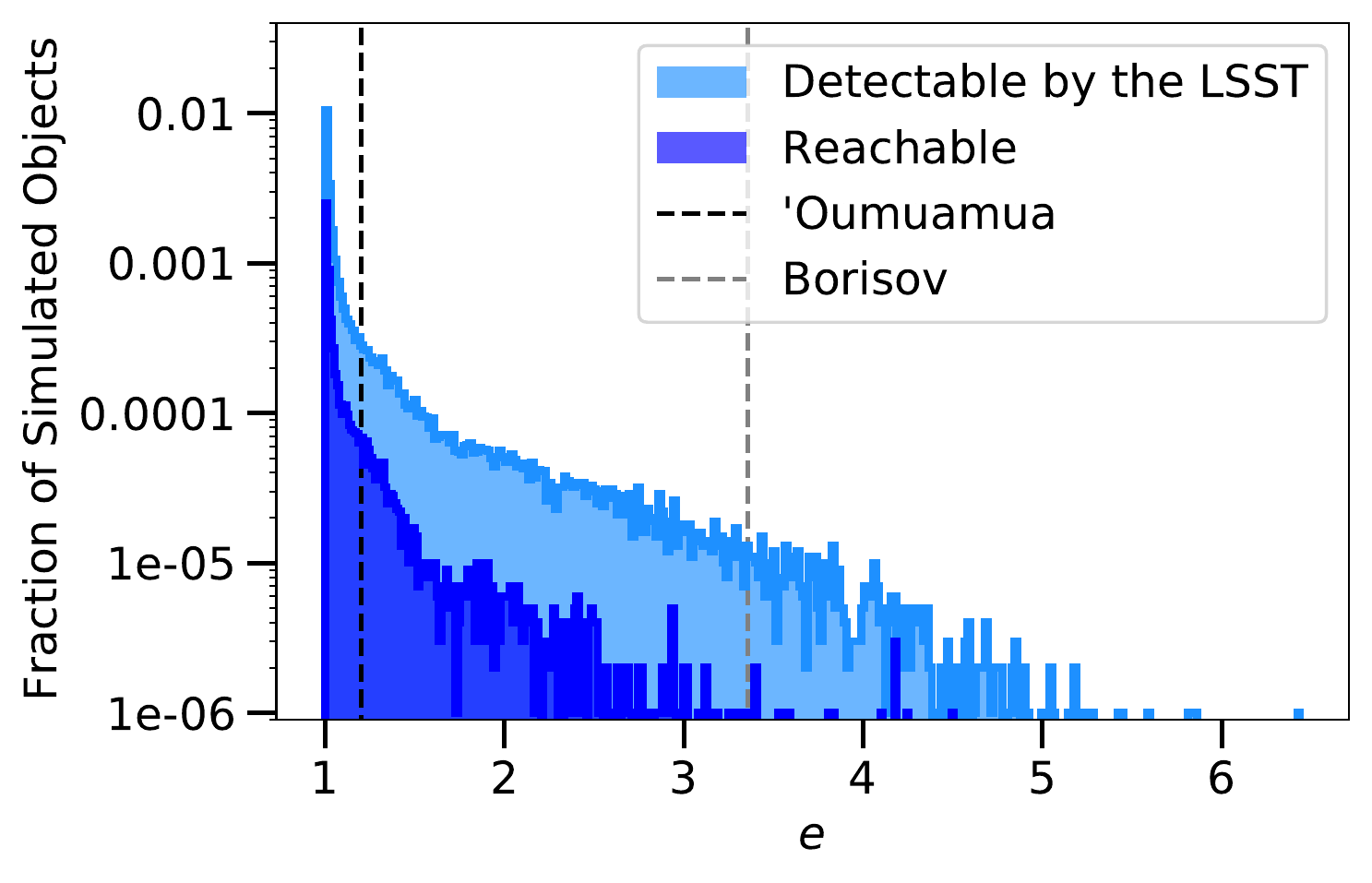}
    \caption{Distributions of eccentricities for the detectable and reachable populations.  The eccentricities of `Oumuamua and Borisov are shown in black and grey dashed lines, respectively.  The typical interstellar object has  eccentricity close to unity.}
    \label{fig:edist}
\end{figure}

\begin{figure}
    \centering
    \includegraphics[scale=0.5,angle=0]{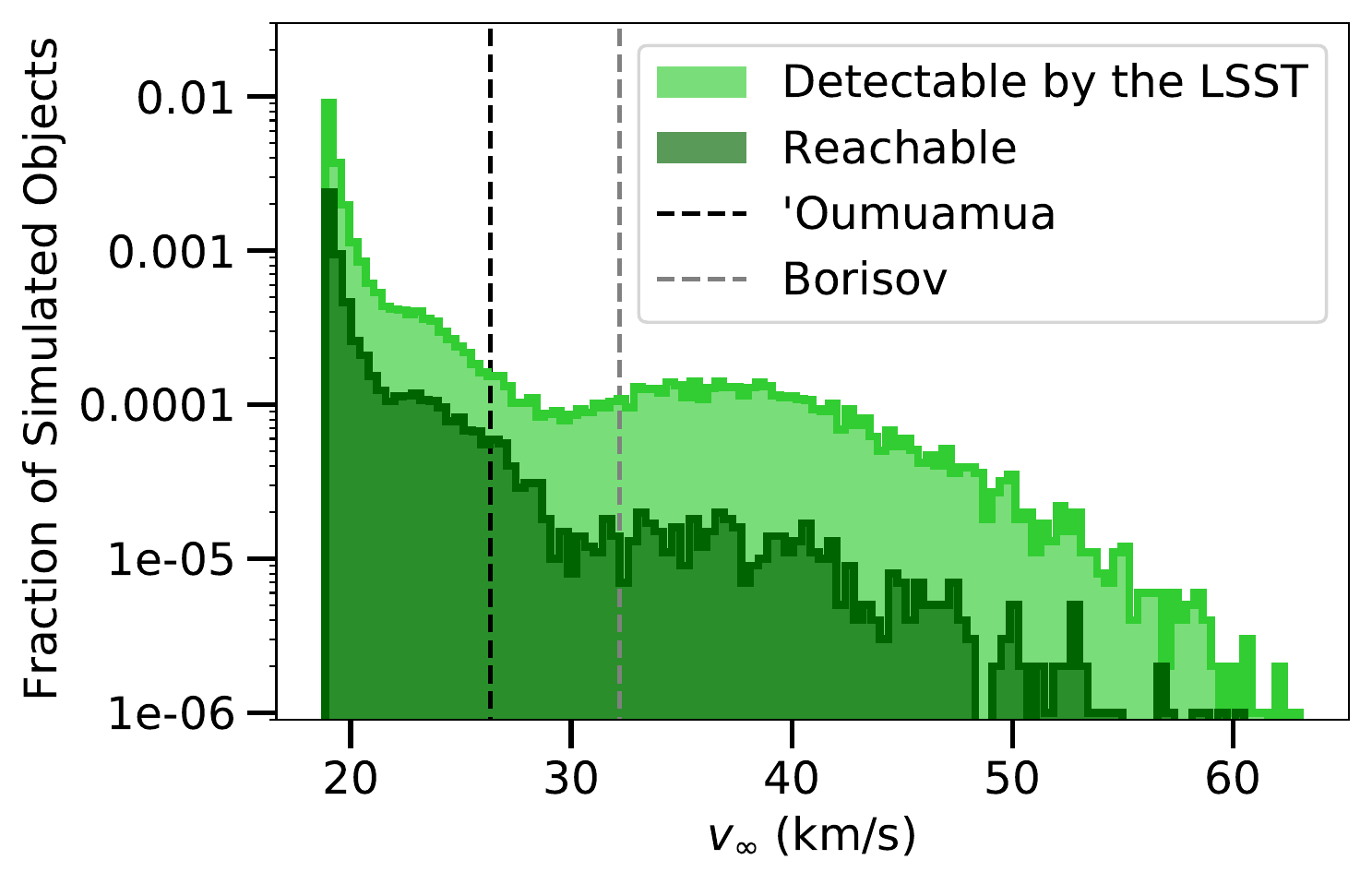}
    \caption{The distribution of hyperbolic excess velocity (incoming velocity) for the detectable and reachable populations. The incoming velocities of `Oumuamua and Borisov are shown in black and grey dashed lines, respectively. }
    \label{fig:vinf}
\end{figure}

In Figure \ref{fig:b_dist}, we show the distribution of impact parameters for detectable and reachable objects. The  limiting magnitude of the LSST creates a sharp cutoff at $b\sim3$ au for objects that will be detectable.  If future interstellar objects have brighter absolute magnitudes than `Oumuamua, this cutoff will increase. These simulations can  easily be scaled to probe the impact parameter cutoff as a function of intrinsic magnitude. In Figure \ref{fig:rmin}, we show the same distributions, but for perihelia, $r_{min}$, instead of impact parameter. In terms of perihelia, the LSST cutoff is at $r_{min}\sim2$au, with this limit being imposed by the detectability criteria of $m\leq24$ for an $H=22.4$ object (as per Figure \ref{fig:min_dist}, all detectable objects must pass within $\sim1.5$ au of the Earth).

In Figures \ref{fig:edist} and \ref{fig:vinf}, we show the distributions of eccentricity and hyperbolic velocity, as in Figure \ref{fig:b_dist}. The distribution of eccentricity is roughly log-normal, while the distribution of hyperbolic velocity has two peaks and significant substructure. The distributions for objects that are reachable for an interception mission roughly mirrors those of the detected population. This implies that the objects that we will be able to probe with an \textit{in-situ} measurement roughly represents the detectable sample for these parameters. 

\begin{table*}
    \centering
    \textbf{H=22.4}\\
    \vspace{0.5em}
    
    \begin{tabular}{||c c c c||}
    \hline
    Criterion & Percent& Conservative Rate per Year&Optimistic Rate per Year\\
    \hline\hline
    $m\leq 24$ & 6.956\% & $\sim 2.3$ & $\sim 4.6$\\
    \hline
    Detectable by the LSST & 2.828\% & $\sim 0.9$ & $\sim 1.9$\\
    \hline
    Detectable, $\Delta v < 30$ km/s & 1.055\% & $\sim 0.35$ & $\sim 0.7$\\
    \hline
    Detectable, $\Delta v < 15$ km/s & 0.424\% & $\sim 0.1$ & $\sim 0.3$\\
    \hline
    Detectable, $\Delta v < 2$ km/s & 0.002\% & $\sim 0.0007$ & $\sim 0.001$\\
    \hline
    \end{tabular}
    
    \vspace{1em}
    
    \centering
    \textbf{H=23.4}\\
    \vspace{0.5em}
    
    \begin{tabular}{||c c c c||}
    \hline
    Criterion & Percent& Conservative Rate per Year&Optimistic Rate per Year\\
    \hline\hline
    $m\leq 24$ & 4.241\% & $\sim 1.4$ & $\sim 2.8$\\
    \hline
    Detectable by the LSST & 1.457\% & $\sim 0.5$ & $\sim 1$\\
    \hline
    Detectable, $\Delta v < 30$ km/s & 0.793\% & $\sim 0.3$ & $\sim 0.5$\\
    \hline
    Detectable, $\Delta v < 15$ km/s & 0.347\% & $\sim 0.1$ & $\sim 0.2$\\
    \hline
    Detectable, $\Delta v < 2$ km/s & 0.002\% & $\sim 0.0007$ & $\sim 0.001$\\
    \hline
    \end{tabular}
    
    \vspace{1em}
    
    \centering
    \textbf{H=24.4}\\
    \vspace{0.5em}
    
    \begin{tabular}{||c c c c||}
    \hline
    Criterion & Percent& Conservative Rate per Year&Optimistic Rate per Year\\
    \hline\hline
    $m\leq 24$ & 1.933\% & $\sim 0.6$ & $\sim 1.3$\\
    \hline
    Detectable by the LSST & 0.378\% & $\sim 0.1$ & $\sim 0.25$\\
    \hline
    Detectable, $\Delta v < 30$ km/s & 0.284\% & $\sim 0.09$ & $\sim 0.2$\\
    \hline
    Detectable, $\Delta v < 15$ km/s & 0.182\% & $\sim 0.06$ & $\sim 0.1$\\
    \hline
    Detectable, $\Delta v < 2$ km/s & 0.002\% & $\sim 0.0007$ & $\sim 0.001$\\
    \hline
    \end{tabular}
    
    \vspace{1em}
    
    \centering
    \textbf{H=25.4}\\
    \vspace{0.5em}
    
    \begin{tabular}{||c c c c||}
    \hline
    Criterion & Percent& Conservative Rate per Year&Optimistic Rate per Year\\
    \hline\hline
    $m\leq 24$ & 0.739\% & $\sim 0.2$ & $\sim 0.5$\\
    \hline
    Detectable by the LSST & 0.099\% & $\sim 0.03$ & $\sim 0.07$\\
    \hline
    Detectable, $\Delta v < 30$ km/s & 0.091\% & $\sim 0.03$ & $\sim 0.06$\\
    \hline
    Detectable, $\Delta v < 15$ km/s & 0.073\% & $\sim 0.02$ & $\sim 0.05$\\
    \hline
    Detectable, $\Delta v < 2$ km/s & 0.001\% & $\sim 0.0003$ & $\sim 0.0007$\\
    \hline
    \end{tabular}
    
    \caption{The annual frequency that interstellar objects are detectable by the LSST and reachable with various $\Delta$v criteria.  \textit{Top panel}: Statistics for our sample of interstellar objects, which assumes that each object's absolute magnitude is equal to that of `Oumuamua.  \textit{Second panel}: Statistics for the same sample after increasing each object's absolute magnitude by 1.  \textit{Third panel}: Statistics for the original sample after increasing each object's absolute magnitude by 2.  \textit{Bottom panel}: Statistics for the original sample after increasing each object's absolute magnitude by 3.  Our percentages are significantly different even if we assume that each object's absolute magnitude is greater than that of `Oumuamua by 1.  The frequency of detectable objects is therefore sensitive to their intrinsic absolute magnitude.}
    \label{percent_table}
\end{table*}

In Table \ref{percent_table}, we show the fractional percentages  of objects that will be detected by the LSST and attainable targets for a range of limiting $\Delta$v.  To convert these fractions to detection rates,  we consider a typical ISO travelling within the 5 au sphere centered at the Sun. We assume a spatial number density of $n_{o}\sim1\times 10^{-1}\,$\textsc{au}$^{-3}$,  lower than that calculated by \citet{Do2018} given the non-detection of an additional `Oumuamua like object since 2017 \citep{Levine2021}. With this assumed spatial density, there are $\sim 50$ interstellar objects within the 5au sphere at any given time.

The crossing time of the 5au sphere varies based on the trajectory and incoming velocity vector of a given interstellar object.  We use incoming velocities as our values for mean crossing velocities, implicitly averaging over the Solar acceleration.  Although the distribution of hyperbolic excess velocity does not decrease monotonically, its frequency is highest at $\sim 20$ km/s and begins to decrease at $\sim 40$ km/s, as demonstrated in Figure \ref{fig:vinf}.  Therefore, we use mean crossing velocities of 20 and 40 km/s to represent our conservative and optimistic predictions for the detection frequency.

The typical distance traveled can be approximated as the average distance, $||d||$, of a chord between two points on a sphere, as described by \citet{Solomon1978}. For a 2-dimensional circle, this may be calculated using the following equation,

\begin{equation}\begin{split}
    ||d||=\frac{1}{(2\pi)^2}\int_0^{2\pi}\int_0^{2\pi}\, \bigg(2-2\cos{\big(\theta_1-\theta_2\big)}\bigg)^{1/2}r\,d\theta_1d\theta_2\,,\end{split}
\end{equation}
which yields $||d||=4/\pi r$. For a sphere, 

\begin{equation}\begin{split}
    ||d||=\frac{1}{(4\pi)^2}\int_0^{2\pi}\int_0^{\pi}\int_0^{2\pi}\int_0^{\pi}\, \bigg(2-2\cos{\phi_1}\cos{\phi_2}\\-2\cos{\theta_1}\cos{\theta_2}\sin{\phi_1}\sin{\phi_2}-2\sin{\theta_1}\sin{\theta_2}\sin{\phi_1}\sin{\phi_2}\bigg)^{1/2}\\r\,\sin{\phi_1}\sin{\phi_2}d\phi_1d\theta_1d\phi_2d\theta_2\,,\end{split}
\end{equation}
which gives $||d||=4/3r$. We verified these values numerically by calculating the average  of sets of points drawn uniformly from  the boundary of the unit circle and sphere.  

It is noteworthy that the average chord length depends on the method by which each chord is constructed (see Table 5 in \citet{Solomon1978}). For the case considered here, where two random points on the surface of the sphere are connected, $||d||=4/3r$, which is also 4 times the volume divided by the surface area. If the chord is generated from a normal to the plane of a random great circle, through a random point in the circle of intersection, $||d||=4/3r$ still. However, if the chords are formed by choosing a single random point on the sphere, and then a random direction vector, $||d||=r$. If the chords are generated by selecting two random points \textit{within} the sphere, then $||d||=12/7r$. For our problem, the method of selecting two random points on the surface is most appropriate. This is a simplified version of the set of problems considered by \citet{alagar_1976,Dunbar1997,burgstaller_pillichshammer_2009}, who calculated similar quantities for higher dimensional cases, such as for a hypersphere $X\subseteq\mathbb{R}^{s}$. Interestingly, this geometric probability problem regarding the distribution of chords within a sphere was developed as an application to cellular biology. Specifically, this formalism was required to quantify the extent with which members of pairs of chromosomes were positioned randomly with respect to other members during mitosis \citep{Barton1963,David1964}.

These values  correspond to crossing times of $\sim 0.8$ and $\sim 1.6$ years. To convert the fractions to detection rates, we use the following equation,

\begin{equation}\begin{split}
    {\rm rate}=33.11\, {\rm yr}^{-1}\bigg(\frac{n_o}{1\times 10^{-1}\,\rm{au}^{-3}}\bigg)\\\bigg(\frac{V}{4/3\pi (5 \rm{au})^3}\bigg)\bigg(\frac{v}{20\,{\rm km/s}}\bigg) \bigg(\frac{(4/3)\,5{\rm au}}{||d||}\bigg)\,, \end{split}
\end{equation}
where $V$ is the simulated volume and v is the assumed typical velocity.
Therefore, we multiply the percentages in Table \ref{percent_table} by $\sim 66$ and $\sim33$ objects to represent the yearly steady state population of the 5au sphere. 

We estimate that the LSST should detect between $0.9-1.9$ interstellar objects like `Oumuamua ($H=22.4$) every year. Therefore, by the end of its 10 year observational campaign, the census of detected interstellar objects should be roughly $9-19$ in magnitude, which is a significant increase from previous estimates. Moreover, with a dedicated mission such as \textit{Bridge}, which should be able to generate a $\Delta$v$\sim 15$km/s, there should be of order $\sim 1-3$ objects detected in the lifetime of the LSST that would be attainable targets. If the \textit{Comet Interceptor} is capable of generating a $\Delta$v$\sim 2$km/s, then it appears that there is a  $\sim 0.07\%$ chance that a target is detected.  

It is important to note that these estimates do not take into account the size-frequency distribution of interstellar objects and the possibility of cometary activity, as discussed previously in this paper. This could provide a significant increase to the number of detected interstellar objects and feasible targets for rendezvous missions.

However, we consider the H-frequency distribution of interstellar objects by running our simulations with the assumption that the absolute magnitudes of each object are higher (fainter) than that of `Oumuamua by 1, 2, and 3.  Table \ref{percent_table} also displays the fractional percentages of objects that meet various criteria under these assumptions.  We find that the percentages are sensitive to our initial choice of absolute magnitude.  For example, while the percentage of objects that reach $m\leq 24$ is 6.956\% if we assume an absolute magnitude of 22.4, the percentage of objects that meet the same criterion decreases by a factor of $\sim 1.6$ if we assume an absolute magnitude of 23.4.  Until an intrinsic absolute magnitude distribution can be determined for ISOs, population estimates will vary significantly depending on the assumptions made.

These detection rates are significantly higher than the ones calculated in \citet{Trilling2017} and \citet{Cook2016}. The rates are still very uncertain, since they are based on the detection of only two objects. Future detections of interstellar objects will refine our estimates of spatial number density in the inner Solar System. These simulations and the resulting percentages in Table \ref{percent_table} can be readily updated when the intrinsic spatial number density is better constrained.  

Because the LSST is capable of surveying the entire sky in a few days, we assumed an ISO detection efficiency  of $100\%$. However, it is likely that the detection efficiency is  lower, which would decrease all of our estimates presented in Table \ref{percent_table}. However, the ability of the LSST to detect transient objects has been demonstrated for the Near Earth Objects, JFCs and LPCs \citep{solontoi2011comet,Veres2017,veres2017b,Jones2018}. \citet{Engelhardt2014} found that there is a $\sim35\%$ chance that a detectable ISO will not be recognized as such due to lack of follow-up observations. However, their investigation involved surveys that have lower ISO detection efficiencies than those of the LSST.  If we adopt $65\%$ as a minimum value for the LSST's ISO detection efficiency, then we still expect that the LSST will detect at least $\sim 6$ ISOs. The detection of these objects will drastically expand our knowledge of this novel field of astrophysics and planetary science.

\section{Conclusions}\label{section:Conclusions}

In this paper, we simulated a population of interstellar objects drawn from their galactic kinematic distribution. Using this synthetic population, we evaluated the distribution of interstellar objects that should be detectable with the forthcoming LSST. We showed that the incoming interstellar objects are strongly clustered in the direction of the solar apex. The distribution of detectable locations on the sky is more isotropic than the distribution of initial locations with slight enhancements in the vicinity of the apex and anti-apex. This is due to the fact that by the time that the objects are detectable by the LSST, they have moved significantly through their trajectories. We showed that less sensitive all sky surveys are more biased to detecting objects close to the ecliptic.

We calculated the distribution of orbital trajectories for the objects that are detected by the LSST. Moreover, we evaluated a subset of these objects that will be reachable for \textit{in-situ} interception missions. We estimated that the LSST should detect of order $\sim 9-19$ interstellar objects every year, and that $\sim 3-7$ reachable targets with a dedicated mission, such as \textit{Bridge} will be detected over the survey's lifetime. There is a $\sim 0.07\%$ chance that a target  for the \textit{Comet Interceptor} will be detected. 

For a larger portion of interstellar objects to be reachable, more efficient technologies, such as solar sails, are required.  \citet{Linares2020} proposed a ``statite'' (static-satellite) concept in which solar sails are utilized to maintain a  probe's orbit and to power an intercept trajectory.  Such a mission could hypothetically achieve $\Delta v\approx 100$ km/s, far higher than the $\Delta$v requirements of \textit{Bridge} or \textit{Comet Interceptor}.  However, these missions require that the orbits of the potential target be known 4-16 months in advance.  Figure \ref{fig:waittime} establishes a tentative upper limit on the wait times of reachable objects of $\sim 5$ months.  Therefore, a statite must be in the optimal position prior to its intercept maneuver.  This type of mission would most likely require a pre-existing cluster of statites in Earth-like heliocentric orbits to be feasible. 

Since a dedicated ISO survey would have virtually the same methodology as a dedicated NEO survey, existing NEO surveys may increase the number of detected ISOs beyond the number of ISOs that we expect the LSST to detect.  Since `Oumuamua and Borisov were detected using Pan-STARRS and the Crimean Astrophysical Observatory \citep{Meech2017, deleon2019}, an international observational campaign may be a worthwhile use of resources.  Dedicated surveys with limiting magnitudes around or fainter than 22, such as Pan-STAARS or ATLAS \citep{Chambers2016, Tonry2018}, may observe an ISO while searching in the vicinity of the ecliptic.  Targeted searches that are equally or more sensitive than the LSST could focus their searches on the vicinity of the solar apex.

\section{Acknowledgements}\label{section:Acknowledgements}
This work utilized the computational resources of the NIH HPC Biowulf cluster (http://hpc.nih.gov).  We thank David Hoover for access to the Biowulf computing cluster.  We thank Adina Feinstein, Amir Siraj, Jacob Bean, Darin Ragozzine, and Ivan Levcovitz for useful conversations and suggestions.  We acknowledge support from the University of Chicago Quad Faculty Research Grant program and Dean's Fund for Undergraduate Research.  We also thank the two anonymous reviewers, whose reports led to modifications that improved the quality of this paper.


\bibliography{sample631}{}
\bibliographystyle{aasjournal}



\end{document}